%% file: main.tex
\documentclass[11pt]{article}
\input{macros.tex}
\date{ }
\begin{document}

\title{Parallel Batch-Dynamic Coreness Decomposition \\ with Worst-Case Guarantees}
\author{Mohsen Ghaffari \\ \small MIT \\ \small ghaffari@mit.edu\and Jaehyun Koo \\ \small MIT \\ \small koosaga@mit.edu}
\maketitle

\begin{abstract} 
We present the first parallel batch-dynamic algorithm for approximating coreness decomposition with worst-case update times. Given any batch of edge insertions and deletions, our algorithm processes all these updates in $\poly(\log n)$ depth, using a worst-case work bound of $b\cdot \poly(\log n)$ where $b$ denotes the batch size. This means the batch gets processed in $\tilde{O}(b/p)$ time, given $p$ processors, which is optimal up to logarithmic factors. Previously, an algorithm with similar guarantees was known by the celebrated work of Liu, Shi, Yu, Dhulipala, and Shun [SPAA'22], but with the caveat of the work bound, and thus the runtime, being only amortized. 

\end{abstract}
\maketitle

\section{Introduction}
This paper presents the first parallel batch-dynamic algorithms for approximating coreness decomposition, and some closely related problems such as (subgraph) density estimation, arboricity estimation, and low out-degree orientations. These algorithms process any batch of edge insertion and deletions in polylogarithmic depth and use work linear in the batch size up to logarithmic factors. Next, we review the related context and state of the art, and then state our results formally.

\subsection{Context: Problems and the Computation Model}
\paragraph{Problems: Coreness decomposition}
Coreness decomposition, sometimes called $k$-core decomposition, is a natural density-based approach to organizing the graph structure. It has a wide range of applications, e.g., in machine learning, databases, social network analysis, and computational biology~\cite{alvarez2005large,esfandiari2018parallel,ghaffari2019improved,bonchi2014core,chu2020finding,esfahani2019efficient,li2020efficient,medya2020game,li2020efficient,dhulipala2017julienne,dhulipala2021theoretically,kabir2017parallel,khaouid2015k,ciaperoni2020relevance,kitsak2010identification,liu2015core,malliaros2016locating}. We will usually denote the main graph in the problem as $G=(V, E)$, and define $n:=|V|$. The coreness $core(v)$ of a node $v\in V$ is the maximal value of $k$ such that there is an induced subgraph $G[S]=(S, E[S])$, with $S\ni v$, that has minimum degree at least $k$. The coreness values provide a hierarchical graph decomposition: Each $k$-core is one connected component of the subgraph induced by nodes with coreness at least $k$. Notice that each $k'$-core, for $k'\geq k+1$, is a subgraph of a $k$-core. 

\paragraph{Problems: Density, Arboricity, and Low Outdegree Orientation} The density $\rho(S)$ of any induced subgraph $G[S]$ for a nonempty subset $S\subseteq V$ is simply $\frac{|E[S]|}{|S|}$ and the density $\rho(G)$ of the graph $G$, or more accurately the density of its densest subgraph, is $\rho(G):=\max_{S\subseteq V} \frac{|E[S]|}{|S|}$. A closely related measure is the arboricity $\lambda(G)$, which is defined as the minimum number of forests to which one can decompose edges of $G$, and by a result of Nash-Williams~\cite{nash1964decomposition} can be defined equivalently as $\lambda(G):=\max_{S\subseteq V, |S|\geq 2} \lceil \frac{|E[S]|}{|S|-1}\rceil$. Notice the close relation: for any graph $G$, we have $\rho(G)\leq \lambda(G) \leq 2\rho(G)$, and if $G$ is simple---i.e., has no self-loops or parallel edges---then we have $\rho(G)\leq \lambda(G) \leq \lceil{\rho(G)+1}\rceil$. In the low out-degree orientation problem, the task is to orient the edges of a given undirected graph $G$ such that we minimize the maximum out-degree. Notice that $\rho(G)$ is a clear lower bound; we usually want to be within a small constant factor of this.

\paragraph{Computational Model: Work and Depth in Parallel Algorithms}  We seek parallel algorithms that solve the above graph problems as the graph undergoes updates. We first recall the parallel model aspects and then discuss the relevant dynamic aspects.

We follow the standard work-depth terminology~\cite{blelloch1996programming} for analyzing the parallelism in algorithms.\footnote{Also, we consider throughout a shared-memory PRAM model, which for simplicity is assumed to be in the strongest PRAM model variant, with concurrent reads and writes (CRCW). The results extend easily to the weaker variants, e.g., with exclusive read and writes (EREW), as the latter can simulate the former with a logarithmic overhead in depth and work.} For an algorithm $\mathcal{A}$, its \textit{work} $W(\mathcal{A})$ is defined as the total number of operations, and its \textit{depth} $D(\mathcal{A})$ is defined as the longest chain of operations with sequential dependencies. Brent's principle~\cite{brent1974parallel}, these bound the time $T_p(\mathcal{A})$ for running the algorithm when we have $p$ processor: $\max\{W(\mathcal{A})/p, D(\mathcal{A})\} \leq T_{p}(\mathcal{A}) \leq W(\mathcal{A})/p + D(\mathcal{A}).$ 

\paragraph{Computational Model: Batch-Dynamic Parallel Algorithms} The area of \textit{batch-dynamic parallel algorithms}---see e.g., \cite{acar2019parallel,acar2020parallel,dhulipala2021parallel,tseng2022parallel,liu2022parallel,ghaffari2023nearly, anderson2024deterministic, ghaffari2024parallel} for some recent work---considers settings where the graph undergoes updates, namely edge insertions and deletions. We want dynamic algorithms that can quickly adjust their solutions to these graph updates, without having to recompute things from scratch, and we particularly would like to leverage parallelism in this. The updates happen in potentially large batches. For a batch of $b$ updates, one clearly has to do at least $b$ work--though often potentially much more. To have an ideal algorithm with perfect parallelism, the best scenario (up to $\log n$ factors) is to process each such batch using $b\poly(\log n)$ work and $\poly(\log n)$ depth: this would mean $\tilde{O}(b/p)$ time with $p$ processors, which is nearly optimal. 

\paragraph{Amortized vs worst-case work bounds} The above work bound guarantee is usually called \textit{worst-case}, to distinguish it from a weaker guarantee called \textit{amortized}. With an amortized work bound, for a large number of batches, the total work performed during these batches should be near-linear in the total size of the batches, but the work per batch is allowed to be much larger than the size of that batch. This implies a good long-term performance, but short-term performance can be slow and bursty. Concretely, sometimes a tiny batch might take a long time, and, if the system has been in use for some time, then the updates can be very slow for a while.

More generally, dynamic algorithms have two kinds of applications: (I) real-world systems, with innate dynamic updates, and (II) static algorithms, where the algorithm designer gradually solves a problem by creating small dynamic updates. Worst-case guarantees are preferred over amortized guarantees in both scenarios (also in both sequential and parallel computations). In scenario (I), real-world systems prefer to process small batches fast, which is guaranteed only by worst-case bounds. In scenario (II), some static algorithms crucially rely on the worst-case bound since they might not revisit the same graph part frequently enough to allow amortization to bring down the cost. For example, see the recent work of Jiang and Yun \cite{jiang2025parallel}, for static parallel vertex-connectivity computation using a worst-case batch-dynamic parallel algorithm; as they note, an amortized algorithm would not work.

\subsection{State of the Art}
\paragraph{Parallel Batch-Dynamic Algorithms for Coreness Decomposition} The most relevant prior work for us, and indeed the original inspiration for our work, is a recent celebrated result of Liu, Shi, Yu, Dhulipala, and Shun~\cite{liu2023parallelbatchdynamicalgorithmskcore}. They gave the first parallel batch-dynamic algorithm for coreness decomposition (and other related problems, which we review later). Their algorithm computes a $(4+\eps)$-approximation of coreness, for any positive constant $\eps>0$, with $\poly(\log n)$ depth per batch and an \textit{amortized work} bound of $b \cdot \poly(\log n)$ where $b$ denotes the batch size. 
The primary objective of our paper is to strengthen this to a \textit{worst-case} work bound---thus avoiding poor short-term performance discussed before. This turns out to be quite challenging from a technical perspective (and thus also interesting, in our view).

\paragraph{Sequential Dynamic Algorithms} Sequential dynamic algorithms normally operate in single edge insertion or deletions (as they can afford to do so). In contrast, trying to do this in the parallel setting despite many updates arriving simultaneously would mean losing the power of parallelism. Sun et al.~\cite{sun2020fully} gave a sequential dynamic algorithm for $4+\eps$ approximation of coreness, with $O(\log^2 n)$ amortized work per single edge update. 

There is a wide range of literature on the other closely related problems in sequential dynamic algorithms. We highlight some here\footnote{This summary prioritizes qualitative aspects and approximation factors over the exact polylogs in the update time.}, especially the state-of-the-art that is contrastable with our results. Bhattacharya et. al.~\cite{bhattacharya2015space} gave a sequential dynamic algorithm for $4+\eps$ approximation of the densest subgraph's density, using $\poly(\log n)$ \textit{amortized} work per update. Sawlani and Wang~\cite{sawlani2020near} improved this to an algorithm with $\poly(\log n)$ \textit{worst-case} work per update for $(1+\eps)$ approximation. Low out-degree orientations have been extensively studied, starting with a classic work of Brodal and Fagerberg~\cite{brodal1999dynamic} which gave a dynamic algorithm that maintains an orientation with outdegree at most $4\lambda$, as long as we always have $\lambda(G)\leq 4\lambda$, using $O(\log n)$ \textit{amortized} work per update. Kopelowitz et al.\cite{kopelowitz2014orienting} gave a dynamic algorithm for orientation with outdegree $O(\lambda + \log n)$ using $O(\alpha \log n)$ worst-case update time. Henzinger et al.~\cite{henzinger2020explicit} showed that an adaptation of the algorithm of Bhattacharya et. al.~\cite{bhattacharya2015space} maintains an orientation with outdegree $O(\lambda(G))$ using $O(\log^2 n)$ amortized update time. 
Chekuri et al. \cite{chekuri2024adaptive} gave a dynamic algorithm that maintains an $O(\alpha(G))$-outdegree orientation using $O(\log^4 n)$ worst-case update time; see their paper for other trade-offs. 
  
\subsection{Our Results}
We present parallel batch-dynamic algorithms, with worst-case work and depth bounds, that compute constant approximations of coreness decomposition, densest subgraph density, arboricity, and low out-degree orientation. The algorithms use $\poly(\log n)$ depth to process each batch of insertions and deletions, a \textit{worst-case} work bound linear in the size of the batch up to a $\poly(\log n)$ factor. Below, we state these results formally. In these statements, as mentioned above, $core(v)$ denotes the coreness of node $v$ in the current graph $G$, $\rho(G)$ denotes the density of the densest subgraph of $G$, and $\lambda(G)$ denotes the arboricity of the graph $G$.

\begin{restatable}{theorem}{kcore}\label{thm:kcore}(\textbf{Coreness Approximation})
    There is a randomized parallel batch-dynamic data structure that maintains a $4+\epsilon$-approximate coreness for any $\epsilon \in (0, 0.1)$. Formally, the algorithm maintains an estimate $core_{ALG}(v)$ such that $core_{ALG}(v) \in [(\frac{1}{2} - \epsilon) core(v), (2 +  \epsilon) core(v)]$, w.h.p. The algorithm takes: 
\begin{itemize}
    \item for initialization from an empty  $n$-vertex graph, $O(\epsilon^{-1} \log n)$ work, and $O(\epsilon^{-1} \log n)$ depth,
    \item for any batch of edge insertions, $O(\epsilon^{-15} \log^9 n)$ work per inserted edge, and $O(\epsilon^{-12} \log^7 n)$ depth for the entire batch,
    \item for any batch of edge deletions, $O(\epsilon^{-13} \log^8 n)$ work per deleted edge, and $O(\epsilon^{-10} \log^6 n)$ depth for the entire batch.
\end{itemize}
\end{restatable}
This gives a counterpart to the celebrated work of Liu et. al.\cite{liu2023parallelbatchdynamicalgorithmskcore}, but with a worst-case work bound instead of their amortized bound. We note that the approximation guarantee of this algorithm exactly matches theirs\footnote{Even though they call this a $(2+\eps)$-approximation, since the multiplicative gap factor between output and target is at most $2+\eps$. We call it a $4+\eps$ approximation, to be more consistent with the nomenclature of approximation algorithms for maximization/minimization optimizations, since once normalized to be above the target, this is a $4+O(\eps)$ approximation. Then this approximation is more naturally comparable with the related problems, e.g., densest subgraph or low out-degree orientation. See, e.g., Bhattacharya et al.\cite{bhattacharya2015space}, where the same guarantee is called a $4+\eps$ approximation.}, and our algorithms are randomized and work with high probability, similar to theirs. 
However, the polylogarithmic factors in our work and depth bounds are considerably higher, which reflects the challenging nature of the worst-case problem (which will be elaborated on in the technical overview). The paper focuses on delivering the first work-efficient algorithm for this problem.
 
\begin{restatable}{theorem}{density}\label{thm:density}(\textbf{Density/Arboricity Approximation and Low Out-Degree Orientation})
    There is a randomized parallel batch-dynamic data structure that maintains a $(1+\epsilon)$-approximate graph density and $(2 + \epsilon)$-approximate low out-degree orientation for any $\epsilon \in (0, 0.1)$. Formally, the algorithm maintains the following w.h.p.:
    \begin{itemize}
        \item an orientation of all edges where $\delta^+(v) \leq (2 + \epsilon) \rho(G)$ for all $v \in V$,
        \item a density estimate $\rho_{ALG}$ such that $\rho_{ALG} \in [(1 - \epsilon) \rho(G), (1 +  \epsilon) \rho(G)]$, and
        \item an arboricity estimate $\lambda_{ALG}$ such that $\lambda_{ALG} \in [(1 - \epsilon) \lambda(G), (2 + \epsilon) \lambda(G)]$.
    \end{itemize}
    The algorithm takes:
\begin{itemize}
    \item for initialization from an empty $n$-vertex graph, $O(\epsilon^{-1} \log n)$ work, and $O(\epsilon^{-1} \log n)$ depth,
    \item for any batch of edge insertions, $O(\epsilon^{-22} \log^9 n)$ work per inserted edge, and $O(\epsilon^{-18} \log^7 n)$ depth for the entire batch,
    \item for any batch of edge deletions, $O(\epsilon^{-19} \log^8 n)$ work per deleted edge, and $O(\epsilon^{-15} \log^6 n)$ depth for the entire batch.
\end{itemize}
\end{restatable}

Liu et al.\cite{liu2023parallelbatchdynamicalgorithmskcore} had an orientation with outdegree upper bounded by $(4+\eps)\lambda(G)$, with amortized work bounds. Besides giving a worst-case work bound, our algorithm also improves the approximation to $2+\eps$ for arboricity-dependent orientation, and $1+\eps$ approximation for density. However, again, the polylogarithmic factors in our work and depth are higher. 


\subsection{Applications for other graph problems}\label{sec:application_withoutproof}
To showcase the usefulness of our primary results, we use our batch-dynamic low outdegree orientation to obtain batch-dynamic parallel algorithms for maximal matching and colorings, with worst-case guarantees. These results are comparable to those of Liu et al., which had amortized work bounds. 
\begin{restatable}{corollary}{matching}\label{thm:matching}(\textbf{Maximal Matching})
    There is a randomized parallel batch-dynamic data structure that, given an upper bound $\rho_{max}$ on the graph density, maintains a maximal matching, with high probability. The algorithm takes:
    \begin{itemize}
    \item for initialization from an empty $n$-vertex graph, $O(1)$ work, and $O(1)$ depth,
    \item for any batch of edge insertions, $O(\rho_{max} + \log^8 n)$ work per inserted edge, and $O(\log^7 n)$ depth for the entire batch,
    \item for any batch of edge deletions, $O(\rho_{max} + \log^7 n)$ work per deleted edge, and  $O(\log^6 n)$ for the entire batch.
        \end{itemize}

\end{restatable}

\begin{restatable}{corollary}{explicit}\label{thm:explicit} (\textbf{Explicit Coloring})
    There is a randomized parallel batch-dynamic data structure that, given an upper bound $\rho_{max}$ on the graph density, maintains an explicit vertex coloring of $O(\rho_{max} \log n)$ colors, with high probability. Formally, the algorithm maintains an assignment of colors $c : U \rightarrow \{1, 2, \ldots, C\}$ where $C \leq O(\rho_{max} \log n)$ and $c(u) \neq c(v)$ for any edge $(u, v)$ in the graph. The algorithm takes:
    \begin{itemize}
        \item for initialization from an empty $n$-vertex graph, $O(1)$ work, and $O(1)$ worst-case depth,
        \item for any batch of edge insertions, $O(\rho_{max} \log^7 n + \log^8 n)$ work per inserted edge, and $O(\log^7 n)$ worst-case depth for the entire batch,
        \item for any batch of edge deletions, $O(\rho_{max} \log^6 n + \log^7 n)$ work per deleted edge, and  $O(\log^6 n)$ worst-case depth for the entire batch.
    \end{itemize}
\end{restatable}

\begin{restatable}{corollary}{implicit}\label{thm:implicit}(\textbf{Implicit Coloring})
    There is a randomized parallel batch-dynamic data structure that maintains an implicit vertex coloring of $O(\rho(G)^2)$ colors, where $\rho(G)$ is the current density of the graph, with high probability. Formally, the algorithm implicitly maintains an assignment of colors $c : U \rightarrow \{1, 2, \ldots, C\}$ where $C \leq O(\rho(G)^2)$ and $c(u) \neq c(v)$ for any edge $(u, v)$ in the graph. Given a query subset $V_{q} \subseteq V$, it can return the colors of these queried vertices. The algorithm takes:
    \begin{itemize}
        \item for initialization from an empty $n$-vertex graph, $O(\log n)$ work, and $O(\log n)$ worst-case depth,
        \item for any batch of edge insertions, $O(\log^9 n)$ work per inserted edge, and $O(\log^7 n)$ worst-case depth for the entire batch,
        \item for any batch of edge deletions, $O(\log^8 n)$ work per deleted edge, and $O(\log^6 n)$ worst-case depth for the entire batch.
        \item for any subset of queries, $O(\rho(G)^5 + \rho(G)^3 \log n \log^* n)$ work for each queried vertex, and $O(\log^*n \log n)$ worst-case depth for the entire subset.
    \end{itemize}
\end{restatable}

\subsection{Technical Overview} We start with a high-level discussion of the challenge of obtaining worst-case work bounds per update (batch), contrasting with amortized bounds. The point is easier to highlight in the simpler context of sequential dynamic algorithms, which process edge insertion/deletions one by one. For simplicity, we consider the problem of orientation with $O(\lambda)$ outdegree, where $\lambda$ is an upper bound of arboricity that we assume to be given.

\paragraph{A simple sequential amortized algorithm} Here is a simple sequential dynamic algorithm with amortized guarantee (this is a rephrasing of Brodal and Fagerberg~\cite{brodal1999dynamic}): keep an orientation with out-degree at most $5\lambda$. Upon deletion of an edge $\{u, v\}$, do nothing. Upon insertion of an edge $\{u,v\}$, orient it arbitrarily, say from $u$ to $v$, and then also do nothing, unless this breaks the outdegree bound for $u$ and makes it reach $5\lambda+1$. If that happens, flip all outgoing edges of $u$ and make them incoming. This may break the outdegree bound for other nodes. Repeat such total flips of all outgoing edges for any new node that breaks the invariant, until there is no such node.~\footnote{One needs some simple bookkeeping to implement this efficiently, which we will ignore in this discussion.} 

A priori, we may think that this process might take many flips (and indeed it would, in the worst case). After discussing an idealized worst-case algorithm, which will be referenced in this analysis, we will argue that $O(\log n)$ amortized flips are enough per update. 

\paragraph{A worst-case sequential algorithm, if we could ignore computations} If we could ignore computational aspects, there is a way to maintain the orientation with outdegree at most $2\lambda$, using at most $O(\log n)$ flips per update, even in the worst case: if after an insertion a node $u$ breaks this invariant and reaches $2\lambda+1$, there must exist a node $w$ within distance $D=\Theta(\log n)$ in the outreach of $u$, such that the outdegree of $w$ is at most $2\lambda-1$. \footnote{Here is a short reasoning. If there is no such $w$, there would be an impossible quick ball expansion. Namely, because the arboricity is at most $\lambda$, one can see that if there no such node $w$, then $|B_{i+1}(u)|/|B_{i}(u)|\geq 2$ for all $i\in [1, \Theta(\log n)]$. Hence, $B_{D}(u)>n$, which is a contradiction. Here, $B_{i}(v)$ is the set of nodes reachable from $u$ within distance $i$.} Flipping the orientation of all edges on a path from $u$ to $w$ fixes the issue. Of course, the computational aspect of finding such a path to a node $w$ is the main challenge in the worst-case algorithm, which we will discuss later.

\paragraph{Analyzing the amortized algorithm} Using the existence of the worst-case flipping scheme, we argue that the amortized algorithm discussed above makes $O(\log n)$ flips amortized per update to keep the looser $5\lambda$ outdegree bound. We use a potential equal to the number of edges oriented differently in the amortized algorithm and the worst-case algorithm from each of the previous paragraphs. Upon every update, the worst-case scheme increases the potential by at most $\Theta(\log n)$. But then, in the amortized algorithm, with every total flip of all outgoing edges of a node, the disagreement reduces by at least $5\lambda - 2\times 2\lambda = \lambda$, and this takes $5\lambda$ flips. Since the potential is always nonnegative, the amortized algorithm makes $O(\log n)$ changes amortized per update.

\paragraph{Actual worst-case algorithms?} In discussing the worst-case algorithm above, we ignored the main challenge: how does one find such a path $u$-to-$w$ to flip, with small computation? The approach in sequential algorithms maintains much more structure in the graph to enable one to find this path fast. Our batch-dynamic parallel algorithm maintains similar structures, as we discuss next.

\paragraph{A sequential worst-case algorithm of Sawlani and Wang} Let us limit ourselves to the setting where $\lambda \in [\log n, 100 \log n]$. Discussing higher or lower arboricities needs adjustment and extra ideas. We discuss maintaining an orientation with outdegree at most $3\lambda$, in worst-case $\poly(\log n)$ time. This is based on a reinterpretation of the sequential algorithm of Sawlani and Wang~\cite{sawlani2020near}. We assign to each vertex $v$ a \textit{height} equal to its current outdegree $\delta^+(v)$, and we maintain the orientation with the crucial invariant that no edge drops in height by more than $1$. Whenever a new edge $(u, v)$ is inserted, we first direct it from $u$ to $v$ such that $\delta^+(u) \leq \delta^+(v)$. Then, we find a maximal path starting from $u$ where the heights along this path are strictly decreasing. Finally, we reverse the found maximal path. This increases the outdegree, and therefore the height, of the end-node of this path. In contrast, the outdegrees of the intermediate nodes remain unchanged. Moreover, the outdegree of the path's start-node $u$ also remains unchanged because it had increased by one due to the newly inserted edge $(u,v)$ and decreased by one due to the reversed path. 

Why is the maximal path short, and in particular, why is the maximum out-degree (i.e., maximum height in this structure) bounded by $3\lambda$? If the outdegree of a node $u$ becomes $3\lambda$, then all nodes within distance $\log n$ in the outreach of $u$ would have outdegree at least $3\lambda-\log n\geq 2\lambda$, by the level invariant. That would mean the graph expands for $\log n$ distance each time by a $2$ factor, which is impossible. Hence, the path length is always at most $O(\log n)$. Finding it is easy, as we extend the path to an out-neighbor on the lower level in each node until that's impossible. To implement this efficiently, one needs some bookkeeping to maintain the outgoing edges of a node going to lower levels. There is a similar scheme for edge deletions, but let us skip that. 

The above showcases how this structure is useful for maintaining low outdegree orientations (in a limited setting of $\lambda$, for now). In our results, we show that similar structures are also good for approximating coreness decomposition. Let us skip that and move toward the core novelty, batch-dynamic parallelism. So far, we have discussed only a single edge update. To move toward the more challenging setting of multiple simultaneous updates, let us slightly rephrase the single update process as a token-dropping game.

\paragraph{Connections to Token Dropping Game}
We view the above as a token-dropping process, where the token represents one increment in the outdegree:
At first, when edge $(u, v)$ was inserted and we oriented it from $u$ to $v$, we see that as initiating a \textit{token} at node $u$. Then, the token will gradually drop in level, step by step, as follows: Whenever the token resides in a node $w$ that has an outgoing edge to a lower-level node (i.e., an out-neighbor with a lower out-degree), the token traverses such an outgoing edge and thus \textit{drops} in level. This procedure is repeated until there are no such outgoing edges, at which point the token stays there and increases the height of the endpoint. Notice that this preserves the invariant that no edges drop the height by more than $1$ since there were no outgoing edges to a lower-level node.  

\paragraph{Parallel Token-Dropping and known results} Our batch-dynamic setting necessitates handling many edge insertions simultaneously. Naturally, we would like to model these as many simultaneous tokens, which gives rise to many challenges. First, naively, one would think that each newly inserted edge $(u,v)$ should again be oriented such that $\delta^+(u) \leq \delta^+(v)$, but this is problematic as the outdegrees of different nodes change simultaneously because of concurrent insertions. Hence, it is not even clear at which endpoint to start the token. Furthermore, we cannot assume that each node contains at most one token---the concurrent insertions can create multiple ``tokens" in one node, and perhaps more crucially, multiple tokens might aggregate in the same node along their drop path. Moreover, each edge can send down only one token during the entire run, and once it is used (which means the edge will be reversed), it should be considered removed. Finally, unlike the previous case where we terminate when the path cannot be extended, the terminating condition needed when modeling the path-finding problem as a multi-token process is more complicated since multiple tokens could be in a vertex. 

We aim to drop all tokens quickly to reach a maximal state where no more drops are necessary, which is quite non-trivial. For instance, the naive approach of simply performing all possible token drops at once can increase the vertex's height considerably and break the invariant (in a way that's hard to recover).

We now would like to point out a highly relevant prior work from the literature of distributed graph algorithms: If we add three restrictions to the above process, the problem becomes essentially equivalent to a \textit{Token-Dropping Game} studied by Brandt et al.~\cite{brandt2021efficientloadbalancingdistributedtoken} for load balancing purposes in the LOCAL model of distributed computing. (1) We assume that the tokens are initially placed such that each node has at most one token, (2) the tokens represent newly inserted edges that are oriented toward the higher outdegree endpoint (or arbitrarily, if equal), (3) the process is not allowed ever to place two tokens at the same node. They gave a distributed algorithm that performs this process in $O(H \Delta^2)$ distributed rounds (comparable to the computation depth in our parallel model). Here, $H$ is the maximum height, and $\Delta$ is the maximum degree of vertices.

Unfortunately, there are several challenges in extending their token-dropping algorithm to our problem. First, their algorithm and proof heavily rely on the assumption that the total degree $\Delta$ of each node is small, whereas in our case, we only have a good bound on the out-degree, and the in-degrees can be very large. Second, we must remove the three restrictions listed above when modeling our path-finding process as token dropping, which requires many ideas. Lastly, their algorithm is for the token-dropping process, which models edge insertions. We need to reverse this process for edge deletions, where tokens start at the bottom and move up, requiring completely new ingredients.

\paragraph{Our Dynamic Algorithm}
As one of the main ingredients in our algorithms, we present a batch-dynamic parallel algorithm for balanced orientation. For that, we address the challenges mentioned above and devise modified algorithms and analysis, both for the parallel token-dropping process and for how we find many paths to manage all edge insertions simultaneously via many iterations of the parallel token-dropping process. 

\begin{itemize}
    \item We present the new proof of the distributed token-dropping algorithm of Brandt et al., which only requires the upper bound on the number of outdegrees. With this proof and several technical details, we can obtain a batch-dynamic parallel algorithm for the Token-Dropping game, assuming that no two tokens are placed at the same node.
    \item We devise a low-depth decomposition algorithm that partitions the set of inserted edges into the \textit{token bundles}. Each token bundle satisfies the desired properties we need for the token-dropping game. 
    \item For the deletion case, we define a new setting of \textit{Token-Pushing Game} and discuss a different algorithm. The algorithm shares the high-level idea with the Token-Dropping Game. However, there are separate technical challenges to overcome, and we need to devise different proofs, algorithms, and data structures.
\end{itemize}

\paragraph{Generalization for other arboricity regimes} 
The work and depth requirement \textit{Token-Dropping Game} depends on the maximum height $H$, which can be superlogarithmic to $n$ if the graph has a higher arboricity. We overcome this inefficiency twofold: First, we carefully avoid the computation in the dense part of the graph by truncating the level structure and maintaining only the part of the invariant in the truncated part. Second, we devise a simple sparsification technique by choosing an appropriate sampling probability and running the algorithm only in the sampled graph. This simple approach works since the graph denseness measures are robust to random sampling, which we prove in this paper. By trying every $O(\frac{\log n}{\epsilon})$ sampling rates of $1, 1 - \epsilon, (1 - \epsilon)^2, \ldots$, we obtain an algorithm that maintains all graph denseness measures unconditionally.

\paragraph{Roadmap}
The next section reviews some preliminaries, including definitions and basic tools. In \Cref{sec:balanced_orientations}, we define the notion of \textit{balanced orientation}, which provides a key structure throughout our algorithms, and we also discuss the relation between (the out-degrees in) balanced orientation and various graph denseness measures, such as graph density, arboricity, and coreness. In \cref{sec:alg_orientation}, we describe batch-dynamic parallel algorithms to maintain a balanced orientation of a graph, subject to certain limitations. In \cref{sec:kcore}, we describe approximation algorithms for the density measures using these batch-dynamic balanced orientation subroutines and the properties presented in this section. In \cref{sec:application}, we describe the application of our data structure for matching and coloring.

\section{Preliminaries}
\label{sec:prelim}

\subsection{Definitions}
For a directed graph, we denote each edge as $(u \rightarrow v)$, meaning that the edge is directed from $u$ to $v$. We use $\delta^+(v)$ to denote its outdegree, and $\delta^-(v)$ to denote its indegree. The degree of a directed graph is the sum of its outdegree and indegree. Moreover, for an undirected graph $G = (V, E)$ and a subset of vertices $S \subseteq V$, the \textit{induced subgraph} $G[S] = (S, E[S])$ is a graph with vertex set $S$, such that $E[S] \subseteq E$ is the set of edges in $E$ where both endpoints belong to $S$. For a set $S$, we denote by $Sym(S)$ the collection of all permutations of elements of $S$.

Below, we state the definitions of \textit{coreness, density, arboricity}, which are the graph denseness measures we aim to compute. 

\begin{definition}
    For a permutation $p \in Sym(S)$ over a set $S$, we say $u <_p v$ for $u, v \in S$ if $u$ occurs before $v$ in $p$. We define $u \leq_p v, u \geq_p v, u >_p v$ similarly.
\end{definition}
\begin{definition}[\textbf{Coreness}]\label{def:corenumber}
    For an undirected graph $G = (V, E)$ and a vertex $v \in V$, the \textit{coreness} $core(G, v)$ (or $core(v)$ if the graph is contextually clear) is defined as a minimum $\lambda$ where the following holds: There exists a permutation $p \in Sym(V)$ such that, for each vertex $w \leq_p v$, there exists at most $\lambda$ edges $(u, w) \in E$ such that $u >_p w$. Here, the vertices are compared with respect to their position in the permutation.
\end{definition}

\begin{definition}[\textbf{Arboricity}]
    Given a loopless undirected graph $G = (V, E)$, its \textit{arboricity} $\lambda(G)$ is the smallest $k$ such that there is a partition $E_1 \cup E_2 \cup \ldots \cup E_k = E$ such that each $(V, E_i)$ is a forest. 
\end{definition}

\begin{definition}[\textbf{Density}]
    Given an undirected graph $G = (V, E)$, the \textit{densest subgraph} is a nonempty subset of vertices $\emptyset \subsetneq S \subseteq V$ that maximizes $\rho(S) = \frac{|E(S)|}{|S|}$, and the \textit{density of the graph} is defined as $\rho(G) = \max_{\emptyset \subsetneq S \subseteq V} \rho(S)$.
\end{definition}

\paragraph{Note} The definitions above for notions like arboricity, densest subgraph, etc, are provided for undirected graphs.
Throughout the paper, we sometimes invoke these notions on directed graphs (concretely, on directed graphs resulting from certain orientations of the input undirected graph). In such cases, the precise definition is to ignore the direction of edges and invoke the notion in the undirected version of the graph. 

\paragraph{Equivalent definitions} We use the following two equivalence statements throughout the paper. The former is well-known by a work of Nash-Williams.

\begin{lemma}[Nash-Williams \cite{nash1964decomposition}]\label{thm:nw}
    For any undirected graph $G = (V, E)$ with $|V| \geq 2$ and without self-loops, we have $\lambda(G) = \max_{S \subseteq V, |S| \geq 2} \lceil \frac{|E[S]|}{|S| - 1} \rceil$.
\end{lemma}

\begin{lemma}\label{lem:coreequiv}
    For an undirected graph $G = (V, E)$, the coreness $core(v)$ of a vertex $v \in V$ is defined as the following: $core(v) = \max_{\{v\} \subseteq S \subseteq V} (mindeg(G[S]))$, where $mindeg(H)$ is the minimum degree of a graph $H$. 
\end{lemma}
\begin{proof}
    Let $\lambda = core(v)$. First, for the sake of contradiction, suppose that $core(v) > \max_{\{v\} \subseteq S \subseteq V} (mindeg(G[S]))$. We start with an empty permutation and repeat the following: Select any vertex with a degree less than $\lambda$, add it to the back of the permutation, and delete it from the graph. At the end of the procedure, the partial permutation contains the vertex $v$, as otherwise, there is a vertex with a degree less than $\lambda$ in the remaining graph due to our assumption. We complete the permutation by adding the remaining vertices at the end of the permutation in an arbitrary order. Let $p$ be this permutation. We can see that each vertex $w \leq_p v$ are the ones obtained by adding a vertex with degree less than $\lambda$ - as a result, for all $w$, there exists at most $(\lambda - 1)$ edges $(u, w) \in E$ such that $u >_p v$, which means $core(v) \leq \lambda - 1$ leading to the contradiction.

    Suppose that $core(v) < \max_{\{v\} \subseteq S \subseteq V} (mindeg(G[S]))$, for the sake of contradiction in the opposite direction. Consider a set $S$ with the minimum degree in the induced subgraph $G[S]$ being at least $\lambda + 1$. Let $p$ be the permutation that gives the value $core(v)$. Pick the minimum vertex $w \in S$ per the ordering of $p$. We have $w \leq_p v$, and there are at least $\lambda + 1$ edges $(u, w) \in E$, for $u\in S$, such that $u >_p w$-- these are all edges incident to $w$ in $G[S]$. This is in contradiction with $p$ being the permutation that assigns coreness value $core(v)=\lambda$ to $v$.
\end{proof}

\subsection{Basic tools}
\paragraph{Binary Search Tree and Sorting} We use the parallel red-black tree in \cite{PARK2001415} to deterministically maintain an ordered list. In CRCW PRAM, the algorithm takes $O(\log n)$ work per element and $O(\log n)$ depth in each batch operation. This result implies an $O(\log n)$-depth and work parallel sorting algorithm.

\paragraph{Hash Tables} We use the parallel hash table in \cite{gil1991towards} for maintaining a dictionary. In CRCW PRAM, the algorithm takes $O(1)$ work per element and $O(\log^* n)$ depth in each batch operation.

Whenever our algorithm is deterministic in other parts, we use the binary search tree, as it is deterministic. In some other algorithms we present, which are randomized (notably, the algorithms of \cref{sec:application_withoutproof}), we switch to a hash table instead, to reduce the bounds by a logarithmic factor.

\paragraph{Concentrations} In analyzing our randomized algorithms, we frequently use the Chernoff bound:
\begin{theorem}[Chernoff's bound]\label{thm:chernoff}
For independent Bernoulli variables $X_1, \ldots, X_n$, let $X = \sum_{i = 1}^n X_i$ and $\mu = E[X]$. Then, for any $\epsilon>0$ we have $Pr[X \geq (1+\epsilon) \mu] \leq \exp(\frac{-\epsilon^2 \mu}{3})$, and for any $\epsilon \in (0, 1)$ we have $Pr[X \leq (1 - \epsilon) \mu] \leq \exp(\frac{-\epsilon^2 \mu}{2})$.
\end{theorem}


\section{Balanced Orientations and Relations with Density Measures}\label{sec:balanced_orientations}
Let us start with the definition of a balanced orientation. We then see how such an orientation relates to various problems we want to solve. 

\begin{definition} [\textbf{Balanced and $H$-Balanced Directed Graphs (or Orientations)}] For a directed graph $G = (V, E)$, we call it \textit{balanced} if for each edge $(u \rightarrow v) \in E$, we have $\delta^+(u) \leq \delta^+(v) + 1$, and \textit{$H$-balanced} if for each edge $(u \rightarrow v) \in E$, we have $\min(\delta^+(u), H) \leq \min(\delta^+(v), H) + 1$. An orientation of an undirected graph is called balanced (or $H$-balanced) if the corresponding directed graph is balanced (or $H$-balanced).
\end{definition}

\subsection{Densest subgraph and Arboricity}
We first state the relation between the maximum degree in the balanced orientation and the densest subgraph and arboricity.

\begin{lemma}\label{lem:density}
    For any balanced directed graph $G = (V, E)$ and a parameter $\epsilon \in (0, 0.1)$, we have\\$\max_{v \in V} \delta^+(v) \in [\rho(G), (1 + \frac{\epsilon}{2}) \rho(G) + \frac{4 \log n}{\epsilon}]$. In particular, when $\rho(G) \geq \frac{8 \log n}{\epsilon^2}$, we have $\max_{v \in V} \delta^+(v) \in [\rho(G), (1 + \epsilon) \rho(G)]$.
\end{lemma}

\begin{proof}
    For the lower bound, let $S \subseteq V$ be a set where $\rho(S) = \rho(G)$. Since each edge in the induced subgraph contributes to the outdegree of some vertex $v \in S$ by one, at least one vertex $v \in S$ should have $\delta^+(v) \geq \rho(G) = \frac{|E[S]}{|S|}$.
    
    We prove the upper bound. Given a set $S \subseteq U$, the \textit{expansion} returns a set $S^\prime = S \cup \{v | (u \rightarrow v) \in E, u \in S\}$. Suppose we have a vertex $v$ such that $\delta^+(v) > (1 + \frac{\epsilon}{2}) \rho(G) + \frac{4 \log n}{\epsilon}$. We start from a singleton $S = \{v\}$ and replace it with its expansion for $\frac{4 \log n}{\epsilon}$ times. Within this procedure, all vertices $w \in S$ has $\delta^+(w) \geq \delta^+(v) - \frac{4 \log n}{\epsilon} > (1 + \frac{\epsilon}{2}) \rho(G)$ by the balancedness condition on edges. From this, and by the definition of $\rho(G)$, we have
    \begin{align*}
        E[S^\prime] > |S| (1 + \frac{\epsilon}{2}) \rho(G) \\
        E[S^\prime] \leq \rho(G) |S^\prime|
    \end{align*}

    Combining both, we have $|S^\prime| \geq (1 + \frac{\epsilon}{2})|S|$ for each expansion, and $(1 + \frac{\epsilon}{2})^{\frac{4\log n}{\epsilon}} > n$, reaching a contradiction.
\end{proof}

Using a classical theorem of Nash-Williams \cite{nash1964decomposition}, we obtain the following relation between maximum out-degree and arboricity:

\begin{corollary}\label{cor:arboricity}
    For any balanced directed graph $G = (V, E)$, a parameter $\epsilon \in (0, 0.1)$ such that graph arboricity satisfies $\lambda(G) \geq \frac{16 \log n}{\epsilon^2}$, we have $\max_{v \in V} \delta^+(v) \in [\frac{1}{2} \lambda(G), (1 + \epsilon) \lambda(G)]$.
\end{corollary}\begin{proof}
    By \cref{thm:nw}, $\lambda(G) = \lceil \max_{S \subseteq V, |S| > 1}\frac{|E[S]}{|S|-1} \rceil$. From this, we have $\rho(G) \leq \lambda(G) \leq 2 \rho(G)$, and the result follows from \cref{lem:density}.
\end{proof}

\subsection{Coreness}
Next, we discuss the result for the \textit{coreness} of each vertex. 

\begin{restatable}{lemma}{maincore}\label{lem:maincore1}
For any $H$-balanced directed graph $G = (V, E)$, parameter $\epsilon \in (0, 0.1)$, and a vertex $v \in V$ with $\delta^+(v) < H - \frac{2 \log n}{\epsilon}$, we have $(\frac{1}{2} - \epsilon) core(v) - \frac{2 \log n}{\epsilon} \leq \delta^+(v)$,
where the coreness is computed from the graph $G$ with orientation removed.
\end{restatable}
\begin{proof}
Suppose there is a vertex $v$ such that $\delta^+(v) < (\frac{1}{2} - \epsilon) core(v) - \frac{2 \log n}{\epsilon}$. By \cref{lem:coreequiv}, we can find a core $U$ such that $\{v\} \subseteq U \subseteq V$ and each vertex in $G[U]$ has degree at least $core(v)$. 

Given a set $S \subseteq U$, the \textit{expansion} returns a set  $S^\prime = \{w \mid u \in S, (w \rightarrow u) \in E, w \in U\}$. We start from a singleton $S = \{v\}$ and replace it with its expansion for $\frac{2 \log n}{\epsilon}$ times. Let $X$ be the number of edges $(w \rightarrow u) \in E$ such that $w \in S^\prime, u \in S$. Within this procedure, all vertices $w \in S$ has $\delta^+(w) \leq \delta^+(v) + \frac{2 \log n}{\epsilon}$ by the balancedness condition on edges, hence $X \leq (\frac{1}{2} - \epsilon) core(v) |S^\prime|$. Also, for each vertex $u \in S$, at least $core(v)$ edges connect it in and out from $u$ to $U$. At most $(\frac{1}{2} - \epsilon) core(v)$ of them are outgoing edges, so there is at least $(\frac{1}{2} +\epsilon) core(v)$ incoming edges from $U$ to $u$. By the definition of expansion, we can see all the incoming vertices are included in $S^\prime$, hence $X \geq (\frac{1}{2} + \epsilon) core(v) |S|$. 

Combining both, we have $|S^\prime| \geq (1+4\epsilon)|S|$, and $|U| \geq (1+4\epsilon)^{\frac{2 \log n}{\epsilon}} > n$, reaching a contradiction.
\end{proof}

\begin{lemma}\label{lem:maincore2}
    For any $H$-balanced directed graph $G = (V, E)$ and parameter $\epsilon \in (0, 0.1)$, we have the following:
    \begin{itemize}
        \item if $\delta^+(v) < H - \frac{2 \log n}{\epsilon}$, then $\delta^+(v) \leq (2 +\epsilon) core(v) + \frac{2 \log n}{\epsilon}$,
        \item if $\delta^+(v) \geq H - \frac{2 \log n}{\epsilon}$, then $H\geq (2 +\epsilon) core(v) + \frac{4 \log n}{\epsilon}$.
    \end{itemize}
    where the coreness is computed from the graph $G$ with orientation removed.
\end{lemma}
\begin{proof}
    Let $p \in Sym(v)$ be a permutation from \cref{def:corenumber}, $v \in V$ be a vertex, and let $U = \{u \mid u \leq_p v\}$. 
    
    Given a set $S \subseteq U$, the \textit{expansion} returns a set $S^\prime = \{w \mid u \in S, (u \rightarrow w) \in E, w <_p u\}$. We start from a singleton $S = \{v\}$ and replace it with its expansion for $\frac{2 \log n}{\epsilon}$ times. Let $X$ be the number of edges $(u \rightarrow w) \in E$ such that $w \in S^\prime, u \in S, w <_p u$. For each vertex $w \in S^\prime$, we have at most $core(v)$ number of edges where $(w, u) \in E, w <_p u$ by \cref{def:corenumber}, hence $X \leq core(v) |S^\prime|$. Also, for each vertex $u \in S$, we have $\delta^+(u) \geq \min(\delta^+(u), H) \geq \min(\delta^+(v), H) - \frac{2 \log n}{\epsilon}$, by balancedness condition on edges. For each $u$, there are most $core(v)$ of edges $(u \rightarrow w)$ have $u <_p w$, hence there are at least $\min(\delta^+(v), H) - \frac{2 \log n}{\epsilon}- core(v)$ edges $(u \rightarrow w)$ with $w <_p u$. From the definition of expansion, we can see that all the outgoing vertices are included in $S^\prime$. Hence we have $X \geq(\min(\delta^+(v), H) - \frac{2 \log n}{\epsilon} - core(v)) |S|$. As a result, we get:
\begin{align*}
 core(v) |S^\prime| \geq (\min(\delta^+(v), H) - \frac{2 \log n}{\epsilon} - core(v)) |S|
\end{align*}
Suppose that we have $\delta^+(v) < H - \frac{2 \log n}{\epsilon}$ and $\delta^+(v) > (2 + \epsilon) core(v) + \frac{2 \log n}{\epsilon}$. Then we have $|S^\prime| \geq (1+\epsilon)|S|$. Alternatively, suppose that we have $\delta^+(v) \geq H - \frac{2 \log n}{\epsilon}$ and  $H > (2 + \epsilon) core(v) + \frac{4 \log n}{\epsilon}$. Again, we have $|S^\prime| \geq (1+\epsilon)|S|$. As a result, for both cases, after $\frac{2\log n}{\epsilon}$ iteration we have $|U| \geq (1+\epsilon)^{\frac{2 \log n}{\epsilon}} > n$.
\end{proof}

\section{Batch-dynamic maintenance of balanced orientation}\label{sec:alg_orientation}

This section discusses an algorithm that maintains an $H$-balanced orientation under batch updates of edge insertion and deletion updates. Our algorithm is deterministic and work-efficient in the worst case. Formally, we prove the following theorem in this section:

\begin{restatable}{theorem}{mainstep1}
\label{thm:balanced-ds}
There is a deterministic parallel batch-dynamic data structure $\textsc{Balanced}(H)$, which, given a parameter $H$, maintains an orientation of the undirected edge set such that its directed counterpart is $H$-balanced. The algorithm takes:
\begin{itemize}
    \item for initialization from an empty graph with $n$ vertices, $O(1)$ work, and $O(1)$ worst-case depth,
    \item for any batch of edge insertions, $O(H^{6} \log n)$ work per inserted edge, and $O(H^{6} \log n)$ worst-case depth for the entire batch,
    \item for any batch of edge deletions, $O(H^{5} \log n)$ work per deleted edge, and $O(H^{5} \log n)$ worst-case depth for the entire batch.
\end{itemize}
\end{restatable}

In \cref{sec:orientation_ds}, we state the data structures we maintain in the algorithm. In \cref{sec:orientation_ins}, we prove \cref{thm:balanced-ds} for the insertion case. In \cref{sec:orientation_del}, we prove  \cref{thm:balanced-ds} for the deletion case, which requires several different ingredients.

\subsection{Data Structures}\label{sec:orientation_ds}
For each vertex, we maintain the set of outgoing edges in a balanced binary search tree (BST) of \cite{PARK2001415}. The BST is only used for efficient set operations, and the order of storing edges is not important; nonetheless, we use a nondecreasing order index. The BST supports bulk insertion and deletion in $O(\log n)$ work and depth. These can be trivially initialized in $O(n)$ time. 

To efficiently implement the decremental queries, we need to find a partition of edges where each vertex has at most one outgoing edge. For this, we define the \textit{rank} of each outgoing edges:

\begin{definition}[\textbf{Rank and Truncated Rank}]
    For each edge $e = (u \rightarrow v)$, the \textit{rank} of the edge $rank(e)$ is the relative position of the edge in the BST of edges outgoing from $u$ labeled from $1, 2, \ldots, n$, if $n$ is the number of outgoing edges from $u$. Similarly, the \textit{truncated rank} of the edge $tr(e) = \min(H + 1, rank(e))$.
\end{definition}

We do not explicitly maintain the rank, but we can compute the rank of each edge by maintaining the subtree size in a BST and following the path toward the root in a bottom-up fashion. 

Additionally, each edge is associated with a \textit{label} that is a nonnegative integer in the range $[0, 3]$. These will be used for bookkeeping purposes to maintain outgoing edges. What exactly goes in the label will be described in \cref{sec:overall_decremental}. The default label for all edges is $0$.

The case of incoming edges is slightly more complex. For each vertex $v$ and each integer $i = 1, 2, \ldots, H + 1$ and each integer $c= 0, 1,2, 3$, we maintain a BST that contains all edges $(u \rightarrow v)$, with \textit{truncated rank} $i$ and label $c$. Unlike the outgoing edges, this BST contains the vertex in order of increasing $\min(H, \delta^+(u))$.

Now, we review the possible operations under this data structure. 

\begin{lemma}\label{lem:adj_rev}
    Given a set of $k \le poly(n)$ edges, along with the new labels that will be applied to these edges after reversal, we can reverse all edges to $\{(v_i \rightarrow u_i)\}_{i = 1}^{k}$ in $O(H \log n)$ depth and work per edge. This procedure will not correct the $\delta^+(v)$ value. 
\end{lemma}
\begin{proof}
    For each edge $(u_i \rightarrow v_i)$, we remove the outgoing edge from the BST of $u_i$ and the incoming edge from the BST of $v_i$. This shifts the truncated rank of at most $H + 1$ outgoing edges in each $u_i$. There are $O(k H)$ such edges in total, and it takes $O(k H \log n)$ work and $O(H \log n)$ depth to obtain the set of edges with truncated rank updated. We update such edges in the respective BST of incoming edges. The algorithm requires $O(H \log n)$ depth and $O(k H \log n)$ work.
\end{proof}

\begin{lemma}\label{lem:adj_update}
    Given a set of $k \le poly(n)$ edges $\{(u_i \rightarrow v_i)\}_{i = 1}^{k}$, we can insert or delete all edges in $O(H \log n)$ depth and work per edge. In case of insertion, assume that new labels are also given. This procedure will not correct the $\delta^+(v)$ value. 
\end{lemma}
\begin{proof}
    We first add or remove them in the BST of outgoing edges, which shifts the truncated rank of at most $H + 1$ outgoing edges in each $u_i$. There are $O(k H)$ such edges in total, and it takes $O(k H \log n)$ work and $O(H \log n)$ depth to obtain the set of edges with truncated rank updated. We update such edges in the respective BST of incoming edges. Finally, we add or remove the given $k$ edges in the BST of incoming edges, as we know the truncated rank of $k$ edges, and the rest of the edges have the correct truncated rank. The algorithm requires $O(H \log n)$ depth and $O(k H \log n)$ work.
\end{proof}

\begin{lemma}\label{lem:lazy_init}
    The data structure can be initialized in $O(1)$ time.
\end{lemma}
\begin{proof}
Instead of initializing all $n$ data structures in the beginning, we use a BST to maintain a map from a vertex to its corresponding data structure. For any queries involving new edge addition, if a vertex of interest has no entries in the BST, we initialize a new data structure and add the mapping between the vertex and the pointer of the data structure in $O(\log n)$ time. As the lemmas mentioned above require at least $O(\log n)$ time for each update, this modification does not change any asymptotic cost.
\end{proof}
\subsection{Incremental Updates}\label{sec:orientation_ins}

\subsubsection{Algorithm for Token Bundles}
A \textit{token bundle} is a set of directed edges that is easier to handle in our incremental algorithm. In this section, we will discuss an algorithm that could only handle an update where every edge addition is assumed to be a \textit{token bundle}, and the algorithm for general case (where we are given an undirected edges without any specific conditions) will be shown later. We first define a token bundle in the incremental updates.

\begin{definition}[\textbf{Token Bundle in Incremental Updates}]\label{def:tokenbundleins}
    A \textit{token bundle} is a set of directed edges $\{u_1 \rightarrow v_1, \ldots, u_k \rightarrow v_k\}$ that satisfies the following conditions:
    \begin{itemize}
    \item $\delta^+(u_i) \leq \delta^+(v_i)$ for all $1 \le i \le k$,
    \item $u_i \neq u_j$ for all $1 \le i < j \le k$.
    \end{itemize}
\end{definition}

In our algorithm, we start by adding all the edges in the token bundle to our $H$-balanced orientations. However, we keep the out-degree the same as before. This results in a discrepancy between the actual out-degree and the out-degree maintained by our data structure, which we aim to resolve using the \textit{token-dropping game} concept.

Let $token(v)$ be the \textit{increased} outdegree for the vertex $v$ throughout the process. Hence, we have $token(u_i) = 1$ in the beginning, and after processing all the updates, the outdegree $v$ will be $\delta^+(v) + token(v)$. We use $\delta^+(v)$ to denote the outdegree \textit{before} processing the current incremental updates.

Right after adding all the edges in the graph, the $H$-balancedness condition is violated if there exists an edge $(u \rightarrow v)$ that satisfies all these four conditions: (a) $\delta^+(u) = \delta^+(v) + 1$, (b) $token(u) = 1$, (c) $token(v) = 0$, and (d) $\delta^+(u) < H$.

To resolve the violation, we repeatedly flip the orientation of such edges. This reduces the outdegree of $u$ by $1$ and increases the outdegree of $v$ by $1$ instead. As a result, we have $token(u) = 0, token(v) = 1$. Note that this reduced the quantity $\sum_{v \in V} token(v) \delta^+(v)$ by one. As the quantity is lower-bounded by $0$, we will eventually reach a situation where we cannot flip such a violated edge, which means we have reached a correct $H$-balanced orientation.

We imagine the condition $token(v) = 1$ as a vertex $v$ containing a single token. Specifically, we say that the vertex $v$ is \textit{occupied} with the token if $token(v) = 1$, and $v$ is \textit{empty} if $token(v) = 0$. Note that the invariant $token(v) \in \{0, 1\}$ holds throughout our entire algorithm: Our definition of \cref{def:tokenbundleins} is engineered in a way so that this invariant is held at the beginning of our algorithm.

\begin{definition}[\textbf{Occupied and Empty Vertices}]\label{def:occupied}
    A vertex $v$ is \textit{occupied} if $token(v) = 1$ and \textit{empty} if $token(v) = 0$. Throughout the algorithm, the vertex is always either occupied or empty. 
\end{definition}

The aforementioned operation of resolving the violation can be imagined as \textit{dropping the token} from $u$ to $v$, where we consider each token as an object and each $\delta^+(v)$ as a \textit{level} (or height) of each vertex. Note that we fix the value $\delta^+(v)$ until the end of the update; hence, the vertex level will not change throughout our operations. A token from a vertex $u$ drops to an empty vertex $v$ with its level strictly smaller (by $1$), and it should do so if such a vertex $v$ exists. 

By dropping the token, the edge where the token is dropped becomes unusable since the edge is flipped, and such a flipped edge can not accommodate another token drop in this bundle due to the outdegree condition. Similarly, an edge in the token bundle can never be flipped since it satisfies $\delta^+(u_i) \leq \delta^+(v_i)$. Hence, only the edges in the original graphs are of consideration, and they will not be involved in a flipping act more than once.

Combining all the discussions above, we can obtain a simple algorithm for resolving the violation, or \textit{dropping the token}: As long as there exists an occupied vertex $u$ with $\delta^+(u) < H$ where it is connected to an empty vertex $v$ with a lower level, flip the edge orientation and move the token to $v$. This naive algorithm is work-efficient as each token corresponds to a directed edge in each update. However, this process is highly sequential.

We solve the problem by dropping the token in several \textit{phases}: Each phase drops a lot of tokens simultaneously without any race condition, and we try to finish the execution of the token bundle within a small number of phases. Let $S$ be the set $v$ with $token(v) = 1$ and has $\delta^+(v) < H$. Each \textbf{phase} repeats the following operation for each vertex $v \in S$ in parallel:

\begin{itemize}
    \item Find any outgoing edge $(v \rightarrow w) \in E$ such that $w \notin S$ with $\delta^+(w) = \delta^+(v) - 1$. This is the vertex we are trying to send down a token.  
    \item \textit{Propose} a token to such vertex $w$. If there are no such vertices, do nothing. 
    \item For each $v \notin S$ that received at least one proposal, accept any of them.
    \item If a proposal is accepted, send a token down.
\end{itemize}

If a phase fails to send down any token, then there are no tokens to send down at all, which means the graph is already $H$-balanced. 

This finishes the description of our algorithm for the token bundles. We prove that the number of phases is polynomial in the height $H$:

\begin{lemma}\label{thm:phasenum}
    The algorithm halts after $O(H^3)$ phases.
\end{lemma}

To prove \cref{thm:phasenum}, we use a strategy similar to Lemma 5 in \cite{brandt2021efficientloadbalancingdistributedtoken}. However, we note that their algorithm relies on the maximum degree being small. In our setting, we only have good upper bounds on the maximum \textit{out}-degree. Thus, we must overcome the challenge resulting from large in-degrees, which requires quite nontrivial adaptations of the algorithm~\cite{brandt2021efficientloadbalancingdistributedtoken}. For this, we need two lemmas.

\begin{restatable}[\textbf{Traversal}]{definition}{phasenuma}\label{def:phasenum1}
    For a token $s$, we define its \textit{traversal} $p_s = (v_1, \ldots, v_d)$ as its path to arrive at its final destination. Hence, the token $s$ was in $v_1$ before any phase and finished in $v_d$ after all phases.
\end{restatable}
\begin{restatable}[\textbf{Extended Traversal}]{definition}{phasenumb}\label{def:phasenum2}
    For a token $s$, we define its \textit{extended traversal} $p^*_s = (v_1, \ldots, v_d, \ldots, v_h)$ where $(v_1, \ldots, v_d)$ is its traversal $p_s$, and for each $d \le i < h$, the last token that left $v_i$ goes to node $v_{i+1}$, and no token left $v_h$. 
\end{restatable}

\begin{definition}[\textbf{Active Vertex in Incremental Updates}]\label{def:phasenum3}
    A vertex $v$ is \textit{active} in phase $t$ if it is occupied before phase $t$, and it has an outgoing edge to a vertex in the lower level that is empty before phase $t$. 
\end{definition}

\begin{lemma}\label{lem:active1}
    If the token $s$ has not arrived yet in $v_d$, there exists a point $1 \le i \le h - 1$ in the extended traversal $p^*_s = (v_1, \ldots, v_d, \ldots, v_h)$ where $v_i$ is occupied but $v_{i+1}$ is not. 
\end{lemma}
\begin{proof}
Suppose not, and assume that token $s$ is in node $v_j$ ($j < d$). We know that $v_j, v_{j+1}, \ldots, v_h$ is all occupied; otherwise, we will find such a node $i$. Then, the following chain happens:
\begin{itemize}
    \item $v_h$ will be occupied forever, as it cannot send the token elsewhere. 
    \item $v_{h-1}$ will be occupied forever. Its last operation is supposed to send something to $v_h$, but this is impossible as $v_h$ is occupied forever, so it's past its last operation.
    \item for any $d \le i \le h - 1$, by induction, $v_i$ will be occupied forever.
\end{itemize}
As a result, $v_d$ will be occupied forever, which means the token cannot be sent, and we reach a contradiction.
\end{proof}

\begin{lemma}\label{lem:active2}
Every vertex $v$ could be active for at most $O(H^2)$ phases.
\end{lemma}
\begin{proof}
If a vertex is active, one of the following happens:
\begin{itemize}
    \item It proposes a token and succeeds.
    \item It proposes a token and does not succeed because the proposed node just received another token.
\end{itemize}
The first case can happen at most $O(H)$ times because it will remove one of its outgoing edges.

The second case will happen at most $O(H^2)$ times. Note that every node can receive at most $O(H)$ tokens since it can send at most that number of tokens with outgoing edges, plus possibly store it by itself. If a node proposed a token through its $i$-th outgoing edge and was rejected because it did not have a token but accepted a fresh new one, then such an event can happen by at most $O(H)$ time for each $i$. Summing this, we obtain $O(H^2)$ different phases where such can happen.
\end{proof}

\begin{proof}[Proof of \cref{thm:phasenum}]   
For any token $s$, we first observe that its extended traversal can have length at most $H$, as the level of the vertices is strictly decreasing. By \cref{lem:active1}, for every phase $t$, there exists a vertex $v_i$ in the extended traversal that is active. By \cref{lem:active2}, a vertex can be active by at most $O(H^2)$ times. Hence, if the algorithm runs more than $O(H^3)$ times, some vertex should be active by more than $O(H^2)$ times, contradicting \cref{lem:active2}.
\end{proof}

\cref{thm:phasenum} concludes the high-level description of our low-span incremental algorithm for the token bundle instance. In the next lemma, we provide technical details on the implementation and the work/depth bounds.

\begin{lemma}\label{thm:insert1}
The incremental algorithm can process each token bundle insertion in $O(H^4 \log n )$ depth and work per edge. 

\end{lemma}
\begin{proof}
    We first discuss the implementation of each phase. For each vertex, we find an outgoing edge that we can propose. As there are at most $H$ outgoing edges, we can do this by scanning all outgoing edges. We collect all proposals given by the vertex as a pair $(w_i, v_i)$, where $w_i$ is the vertex being proposed, and $v_i$ is the vertex that proposed. We sort all proposals in lexicographical order and take the $(w_i, v_i)$ such that $w_{i-1} \neq w_i$. Finally, we process each proposal by moving the tokens and reversing each edge. 

    To find an outgoing edge, we must traverse the BST of outgoing edges to find an outgoing edge, which takes $O(H)$ depth and work per edge. Sorting all proposals takes $O(\log n)$ depth and work per edge. We invoke \cref{lem:adj_rev} to reverse an edge, which takes $O(H \log n)$ depth and work per edge.

    Each simulation of phases takes $O(H \log n)$ depth and work per edge. Using \cref{thm:phasenum}, we obtain $O(H^4\log n)$ depth and work per edge.

    Finally, we need to process an insertion of edges for each token bundle and correct the outdegree of each node. The insertion of edges can be done by invoking \cref{lem:adj_update}, which takes $O(H \log n)$ depth and work per edge, which is negligible. Correcting the outdegree is not trivial: We need to update $\delta^+(v)$ to the correct value and maintain the BST of incoming edges to preserve its sorted order. We remove all outgoing edges from $v$ in the BST of incoming edges, update $\delta^+(v)$, and insert all outgoing edges from $v$. The number of vertices concerned is exactly the number of edges inserted, and all vertices concerned have at most $H$ outgoing edges. Hence, this takes $O(H \log n)$ depth and work per edge, which is negligible.
\end{proof}

\subsubsection{Algorithm for General Case}
If an edge $(u, v)$ satisfies the condition $\min(\delta^+(u), \delta^+(v)) \geq H$, adding an edge in any direction will not break the invariant, and it is not necessary to simulate a \textit{token dropping} procedure. We show an algorithm that repeatedly removes \textit{token bundles} in the given incremental edge set until all the remaining edges satisfy $\min(\delta^+(u), \delta^+(v)) \geq H$, which can be processed directly by manipulating the respective data structures. We later argue that only a few such iterations of removing token bundles are sufficient.

We introduce the procedure \textsc{ExtractTokenBundle}, which takes a set of incremental updates and extracts a token bundle with respective orientations.

\begin{itemize}
    \item For each edge $(u, v) \in E$, \textit{propose} the token to the vertex with smaller outdegree.
    \item For each vertex that received at least one proposal, accept any of them.
    \item For each accepted edge, remove it from the set of updates and direct it so that for an edge $u \rightarrow v$, $u$ is the vertex that accepted the proposal.
    \item Insert the created token bundle by \cref{thm:insert1}.
\end{itemize}

Our goal is to argue that after a small number of \textsc{ExtractTokenBundle}, we will end up with edges where $\min(\delta^+(u), \delta^+(v)) \geq H$ holds. We prove that this is true.

\begin{lemma}\label{lem:insert2}
    After $O(H^2)$ execution of \textsc{ExtractTokenBundle} procedure, all remaining edges in the incremental updates satisfies $\min(\delta^+(u), \delta^+(v)) \geq H$.
\end{lemma}
\begin{proof}
    For each vertex $v$, define $tight(v) = \{(v \rightarrow w) \in E \mid \delta^+(w) = \delta^+(v) - 1\}$, and $\Phi(v) = (H + 1) (\delta^+(v) + 1) - |tight(v)|$. Since $0 \le |tight(v)| \leq \delta^+(v)$, $0 \leq \Phi(v) \leq (H + 1)^2$ for each vertex $v$ with $\delta^+(v) < H$.

    For each edge $(u, v) \in E$, we assume that the token was proposed to $u$ without loss of generality. We argue that if $\delta^+(u) < H$, a single execution of \textsc{ExtractTokenBundle} will increase $\Phi(u)$. To see this, we inspect two cases.

    Consider the case where the vertex $u$ ends up with a token. As a token indicates an increase in outdegree, $\Phi(u)$ will increase by $H + 1$ and could decrease by $H$ by any possible change in $|tight(u)|$, which indicates a net increase.
    
    Consider the case where the vertex $u$ does not end up with a token. Its outdegree stays, so it suffices to prove that $|tight(u)|$ decreases. As a vertex $u$ began occupied and ended up empty, at least one of the tokens had dropped using its edges, which will belong to $tight(u)$. For the vertex $u$ to have any new edge in $tight(u)$, it should be one of the reversed edges or the newly added edge. Both edges satisfy a degree condition $\delta^+(u) \leq \delta^+(v)$ for $u \rightarrow v$, but the RHS never decreases, and the LHS will stay as $u$ did not end up with a token.

    Now, we prove the main statement. Note that, if an edge satisfies $\min(\delta^+(u), \delta^+(v)) \geq H$, then it will satisfy that condition forever as $\delta^+(u)$ is nondecreasing. Suppose that an edge remained $\min(\delta^+(u), \delta^+(v)) < H$ throughout the $2(H + 1)^2 + 3$ execution of \textsc{ExtractTokenBundle}. As we increase $\Phi$ value for one of $u, v$ in each execution, one of $u, v$ has $\Phi$ value increased by at least $(H+1)^2+2$ times. This means that we've once increased $\Phi(v)$ when $\Phi(v) \geq (H+1)^2+1$. This happens only if we have $\min(\delta^+(u), \delta^+(v)) \geq H$, reaching a contradiction.
\end{proof}

Finally, we can show \cref{thm:balanced-ds} for the insertion updates.

\begin{lemma}\label{thm:insert2}
The incremental algorithm can process edge insertion updates in $O(H^6 \log n )$ depth and work per edge.
\end{lemma}
\begin{proof}
    The procedure \textsc{ExtractTokenBundle} is implemented in the following way: We collect all proposals given by the edge as a pair $(w_i, e_i)$, where $w_i$ is the vertex being proposed, and $e_i$ is the vertex that proposed. We sort all proposals in lexicographical order and initialize the token bundles with proposals $(w_i, e_i)$ such that $w_{i-1} \neq w_i$. By initializing, we remove them from the query set and assign the orientation so that the edge points toward the higher-level vertices. Finally, we invoke the algorithm of \cref{thm:insert1} with $O(H^4 \log n)$ depth and work per edge. 

    The depth and work per edge are dominated by the algorithm of \cref{thm:insert1}, which is $O(H^4 \log n)$ work and depth. Executing this for $O(H^2)$ times gives us $O(H^6 \log n)$ work and depth.

    Finally, we are left with edge insertions where $\min(\delta^+(u), \delta^+(v)) \geq H$. We do not have to worry about breaking the $H$-balanced condition for those cases. We invoke the algorithm from \cref{lem:adj_update}, obtaining $O(H \log n)$ depth and work per edge, which is negligible. For correcting the outdegree, as $\min(H, \delta^+(v))$ does not change after the update, we do not need to update the data structures.
\end{proof}

\subsection{Decremental Updates}
\label{sec:orientation_del}
For the decremental updates, we follow a framework similar to the incremental ones, but our execution will be different. Here, every edge is deleted immediately after the updates are received, but the out-degree will not be decreased accordingly. Then, we will slowly decrease the out-degree to match the actual out-degree by moving the \textit{token bundle} upward. The token bundle may not correspond to the actual set of edges: It only represents the difference between the out-degree (level) maintained in our orientation and the actual out-degree.

\subsubsection{Algorithm for Token Bundles}
We again define a \textit{token bundle} similarly with the incremental updates. A \textit{token bundle} implies that exactly one outgoing edge from the vertex $\{u_1, u_2, \ldots, u_k\}$ in the $H$-balanced orientation is removed. However, as said earlier, the difference is that the token bundles are not associated with edges. Still, one can think that the special case here is that the algorithm can only remove a set of edges if every node has at most one outgoing edge removed.

\begin{definition}[\textbf{Token Bundle in Decremental Updates}]
    A \textit{token bundle} is a set of distinct vertices $\{u_1, u_2, \ldots, u_k\}$.
\end{definition} 

Each token represents the \textit{decreased} outdegree for the vertex $v$ throughout the process. Hence, we have $token(u_i) = 1$ in the beginning, and after processing all the updates, the outdegree of vertex $v$ will be $\delta^+(v) - token(v)$. We use $\delta^+(v)$ to denote the outdegree \textit{before} processing the current decremental updates. Again, we imagine the condition $token(v) = 1$ as a vertex $v$ containing a single token, and we will keep the invariant $token(v) \in \{0, 1\}$ throughout our entire algorithm. We use the concept of \textit{occupied} and \textit{empty} as defined in \cref{def:occupied}.

Right after removing all the edges in the graph, the $H$-balancedness condition is violated if there exists an edge $(u \rightarrow v)$ that satisfies all these five conditions: (a) Edge $(u \rightarrow v)$ is not deleted, (b) $\min(H, \delta^+(u)) = \delta^+(v) + 1$, (c) $token(u) = 0$, (d) $token(v) = 1$, and (e) $\delta^+(v) < H$.

To resolve the violation, we repeatedly flip the orientation of such edges. This reduces the outdegree of $u$ by $1$ and increases the outdegree of $v$ by $1$ instead. As a result, we have $token(u) = 1, token(v) = 0$. Note that, this increased the quantity $\sum_{v \in V} token(v) \min(H, \delta^+(v))$ by $1$. As the quantity is upper-bounded by $n H$, we will eventually reach a situation where we cannot flip such a violated edge, which means we have reached a correct $H$-balanced orientation.

The aforementioned operation of resolving the violation can be imagined of as \textit{pushing the token} from $v$ to $u$, where we consider each token as an object and $\min(H, \delta^+(v))$ as a \textit{level} (or height) of each vertex. Note that the level of all vertices will not change throughout our operations. A token from a vertex $v$ is pushed to an empty vertex $u$ with its level strictly larger (by $1$), and it should do so if such a vertex $v$ exists. 

By pushing the token, the edge where it is pushing becomes unusable since it is flipped. Due to the outdegree condition, such a flipped edge can not accommodate another token drop in this bundle. 

Combining all the discussions above, we can obtain a simple algorithm for resolving the violation, or \textit{pushing the token}: As long as there exists an occupied vertex $v$ with $\delta^+(v) < H$ where it is connected to an empty vertex $v$ with a high level, flip the edge orientation and move the token to $u$. This naive algorithm is work-efficient as each token corresponds to a directed edge in each update. However, this process is highly sequential.

We solve the problem by pushing the token in several \textit{phases}: Each phase pushes a large number of tokens simultaneously without any race condition, and we try to finish the execution of the token bundle within a small number of phases.

Let $S$ be the set of vertices $v$ with $token(v) = 1$ that have $\delta^+(v) < H$. In each \textbf{phase}, for each vertex $v \in S$, we repeat the following for $i = 1, 2, \ldots, H$:

\begin{itemize}
    \item For each $v \in S$ with a token, find an incoming edge $(w \rightarrow v)$ such that $w \notin S$, $w$ does not have a token, and $\min(H, \delta^+(w)) = \min(H, \delta^+(v)) + 1$, and $rank(w \rightarrow v) = i$.
    \item Send a token along that edge. 
\end{itemize}

We note that the $S$ is not changed after each $i$: $S$ will be updated only after the phase is finished. As a result, a node in $S$ may not have a token, and a node not in $S$ may have a token. We also remind each node has at most one token after the operation, since for $1 \le i \le H$, each node has only one outgoing edge with rank $i$.

Finally, we need to deal with the edges with a rank of at least $H + 1$, or, equivalently, with a truncated rank of exactly $H + 1$. We do the following:

\begin{itemize}
    \item For each $v \in S$ with a token and $\delta^+(v) = H - 1$, find an incoming edge $(w \rightarrow v)$ such that $tr(w \rightarrow v) = H+1$. This implies $w \notin S, \min(H, \delta^+(w)) = \min(H, \delta^+(v)) + 1$, but does not imply that $w$ have no tokens.
    \item Send a token along that edge. 
\end{itemize}

Note that for the case of $tr(w\rightarrow v) = H + 1$, each vertex can receive more than one token - specifically, it can receive multiple tokens from the edges of $tr(e) = H + 1$, or it can receive a token from the edge $rank(e) \geq H$ and then receive at least one token from the edge $tr(e) = H + 1$. Here, we say that any token that passed through the edge $tr(e) = H + 1$ became \textit{transparent}. Any vertex that received a \textit{transparent} token does not count as a vertex that received a token. Hence, a vertex may have a normal token and receive a transparent token, and a vertex with a transparent token may receive a normal token as long as it has not had a normal token before. The rationale for this decision is that a vertex with $k$ outgoing edges of $tr(e) = H + 1$ can accommodate $k$ tokens, as we have $\delta^+(v) = H + k$, and removing $k$ outgoing edge does not change $\min(\delta^+(v), H)$. 

A cleaner way to think about this procedure is that for each edge with $tr(w \rightarrow v) = H + 1$, we replace $w$ to a dummy node of a single outgoing edge without tokens, so any token that passed through $w \rightarrow v$ will not affect the vertex $w$. We use this interpretation to prove the upper bound of the required number of phases.

This finishes the description of each phase. If a phase fails to send down any token, then there is no token to send down, which means the graph is already $H$-balanced. 

We prove that the number of phases is polynomial in the height $H$:

\begin{lemma}\label{thm:phasenum2}
    The algorithm halts after $O(H^3)$ phases.
\end{lemma}

The proof strategy of \cref{thm:phasenum} does not work directly since the number of incoming edges is large. Still, our modification using the \textit{truncated ranks} makes it sufficient to apply the proof of \cref{thm:phasenum} with small modifications. 

We first repeat the definition of \cref{def:phasenum1}, \cref{def:phasenum2}.

\phasenuma*
\phasenumb*

We emphasize that the definition of \textit{active vertex} here differs from the definition we used in incremental updates. The underlining below highlights the difference, and we discuss it again in the follow-up remark.
\begin{definition}[\textbf{Active Vertex in Decremental Updates}]
    A vertex $v$ is \textit{active} in phase $t$ if it is \underline{empty} before phase $t$, and it has an outgoing edge to a vertex in the lower level that is \underline{occupied} before phase $t$. 
\end{definition}
\begin{remark}\label{rem:dummy}
The difference with \cref{def:phasenum3} is underlined. Note that these definitions assume that each vertex has zero or one token, which is not compatible with our transparent token concept. We use the interpretation discussed earlier: For each edge with $tr(w \rightarrow v) = H + 1$, we replace $w$ with a dummy node of a single outgoing edge without tokens in level $H$. With this interpretation, we can still correctly model our algorithm's behavior without modifying any previous assumptions about tokens.
\end{remark}

The following lemma corresponds to the \cref{lem:active1} and \cref{lem:active2} in the incremental case and is crucial in bounding the number of phases.

\begin{lemma}\label{lem:active3}
    If the token $s$ has not arrived yet in $v_d$, there exists a point $1 \le i \le h - 1$ in the extended traversal $p^*_s = (v_1, \ldots, v_d, \ldots, v_h)$ where $v_i$ is occupied but $v_{i+1}$ is not. 
\end{lemma}
\begin{proof}
Suppose not, and assume that token $s$ is in node $v_j$ ($j < d$). We know that $v_j, v_{j+1}, \ldots, v_h$ is all occupied; otherwise, we will find such node $i$. Then, note the following chain happens:

\begin{itemize}
    \item $v_h$ will be occupied forever, as it cannot send the token elsewhere.
    \item $v_{h-1}$ will be occupied forever. Its last operation is supposed to send something to $v_h$, but this is impossible as $v_h$ is occupied forever, so it's past its last operation.
    \item For any $d \le i \le h - 1$, by induction, $v_i$ will be occupied forever.
\end{itemize}
As a result, $v_d$ will be occupied forever, which means the token cannot be sent, and we reach a contradiction.
\end{proof}

\begin{lemma}\label{lem:active4}
Every vertex $v$ could be active for at most $O(H^2)$ phases.
\end{lemma}
\begin{proof}
If a vertex is active, one of the following happens:
\begin{itemize}
    \item It receives a token.
    \item It does not receive a token because all outgoing vertices with a token send the token somewhere else.
\end{itemize}
The first case can happen at most $O(H)$ times, because it will remove one of its outgoing edges. Note that, by \cref{rem:dummy}, every node has at most $H$ outgoing edges.

The second case will happen at most $O(H^2)$ times. Note that every vertex can send at most $O(H)$ tokens since every vertex has at most $H$ outgoing edges to receive one, plus the initial token. As a result, each time a second case happens, one of the outgoing vertices will send a token, which can happen at most $O(H^2)$ times in total.
\end{proof}

\begin{proof}[Proof of \cref{thm:phasenum2}]   
For any token $s$, we first observe that its extended traversal can have length at most $H$, as the level of the vertices is strictly increasing. By \cref{lem:active3}, for every phase $t$, there exists a vertex $v_i$ in the extended traversal that is active. By \cref{lem:active4}, a vertex can be active at most $O(H^2)$ times. Hence, if the algorithm runs more than $O(H^3)$ times, some vertex should be active by more than $O(H^2)$ times, contradicting \cref{lem:active4}.
\end{proof}

\cref{thm:phasenum2} concludes the high-level description of our low-span decremental algorithm for the token bundle instance. In the next lemma, we provide technical details on the implementation and the work/depth bounds.

\begin{lemma}\label{thm:delete1}
The decremental algorithm can process each token bundle deletion in $O(H^{4} \log n)$ depth and work per edge. 

\end{lemma}
\begin{proof}
    Let $k$ be the number of edges to delete.
    
    We first discuss the implementation of each phase. For each edge $w \rightarrow v$, we assign a label $label(w \rightarrow v) = 2 \cdot \mathbf{1}[w \in S] + \mathbf{1}[w \text{ contains a token}]$. Initially, every edge contains a label $0$, so we iterate through all the outgoing edges of $S$ with $rank(w \rightarrow v) \leq H$ and update the label accordingly. As $|S| = k$, we need to update the label of at most $k H$ edges by updating the BST of incoming edges - no truncated rank will change in this procedure.

    Consider the inner loop with $1 \le i \le H$. For each vertex $v \in S$ with a token, we find an incoming edge $(w \rightarrow v)$ such that 
    \begin{itemize}
        \item $w \notin S$, $w$ does not contain a token,
        \item $\min(H, \delta^+(w)) = \min(H, \delta^+(v)) + 1$,
        \item $rank(w \rightarrow v) = i$.
    \end{itemize}
    The first condition states that the label should be $0$. Hence, there is a single BST corresponding to all of these conditions except the second. Since the orientation is $H$-balanced before the updates, and our BST maintains all incoming vertices in the increasing order of $\min(H, \delta^+(w))$, we only have to test the leftmost node in our BST. If the BST is nonempty and the leftmost node satisfies the second condition, we send a token along the edge. After sending a token in each $i$, we update the label for the outgoing edges of the node that received a token. This requires an update on the BST of incoming edges. Processing the edges of truncated rank $H + 1$ is not different since all the labels will be $0$. After finishing all phases, we reverse all edges using \cref{lem:adj_rev} and clear out all the labels to $0$. 
    
    Let's check the depth and work per edge. The initial setting of labels takes $O(\log n)$ depth and $O(k H \log n)$ total work. For each $i$, finding an incoming edge takes $O(\log n)$ depth and $O(k\log n)$ total work. For each $i$, sending a token takes $O(\log n)$ depth and $O(H \log n)$ work per each token. We send at most $k$ tokens, resulting in $O(H \log n)$ depth and $O(k H \log n)$ total work after processing for all $i$. Finally, we reset the labels in $O(\log n)$ depth and $O(k H \log n)$ total work and reverse all edges using \cref{lem:adj_rev}, which takes $O(H \log n)$ depth and $O(k H \log n)$ total work. Each phase simulation takes $O(H \log n)$ depth and work per edge. Using \cref{thm:phasenum2}, we obtain $O(H^4\log n)$ depth and work per edge.
    
    Finally, we discuss the deletion of edges and the correction of the outdegree of each node. Deleting edges can be done by invoking \cref{lem:adj_update} in the beginning, which takes $O(H \log n)$ depth and work per edge, which is negligible. To correct the outdegree, we need to update $\delta^+(v)$ to the correct value and maintain the BST of incoming edges to preserve its sorted order. We remove all outgoing edges from $v$ in the BST of incoming edges, update $\delta^+(v)$, and insert all outgoing edges from $v$. The number of vertices concerned is at most $k$, and each of them has at most $H$ outgoing edges (otherwise, $\min(H, \delta^+(v))$ stays in $H$). Hence, this takes $O(H \log n)$ depth and work per edge, which is negligible.
\end{proof}

\subsubsection{Algorithm for General Case}\label{sec:overall_decremental}
Note that any edge $u \rightarrow v$ where $\delta^+(u) > H$ can be deleted easily without violating the $H$-balancedness condition. We first delete such edges. For this, we sort all the edges $\{(u_i \rightarrow v_i)\}_{i = 1}^{k}$ in the increasing order of $u_i$, and for each vertex $u$ with $\delta^+(u) > H$, we delete the first $\delta^+(u) - H$ edges heading out from vertex $u$. By \cref{lem:adj_update}, this takes $O(H \log n)$ work and depth per each edge, and we need no updates in the data structure as $\min(H, \delta^+(u))$ is unchanged. 

We can now assume $\delta^+(u_i) \leq H$ for all edges $u_i \rightarrow v_i$. For each of the edges, we delete them using \cref{lem:adj_update}, and we add a token in a vertex $u_i$ instead of adjusting the outdegree immediately. As a result, each vertex will contain at most $H$ tokens. Then, we process each token one by one: We establish at most $H$ token bundles containing the partition of our tokens so that every vertex contains at most one token. Finally, we delete the token bundle with \cref{thm:delete1}. 

\begin{lemma}\label{thm:delete2}
The decremental algorithm can process edge deletion updates in $O(H^{5} \log n)$ depth and work per edge.
\end{lemma}
\begin{proof}
    We sequentially invoke the algorithm of \cref{thm:delete1}, which takes $O(H^{4} \log n)$ depth and work per edge. Executing this for $O(H)$ times gives us $O(H^{5} \log n)$ depth and work per edge.
\end{proof}


\section{Batch-dynamic Coreness Decomposition, and Other Problems}\label{sec:kcore}
In this section, we prove the following results:
\kcore*
\density*

\paragraph{Roadmap} We describe the proof of the result step-by-step. In \cref{sec:kcore_fixed}, we prove the main result under a simplifying assumption that the graph's arboricity (or other measures, in case of other result statements) is always in the range $[0.1H, H]$, for a given parameter $H$. Then, in \cref{sec:kcore_all}, we extend the proof so that the assumption is unnecessary.

\subsection{The algorithms, assuming a good arboricity upper bound H}\label{sec:kcore_fixed}
In this subsection, we prove the following results:

\begin{theorem}\label{thm:kcore-fixed}
    There is a randomized parallel batch-dynamic data structure, which, given a parameter $H \geq 1$ and $\epsilon \in (0, 0.1)$, maintains an estimate $f(v)$ which satisfies the following conditions w.h.p.:
    \begin{itemize}
        \item if $f(v) < H$, then $f(v) \in [(\frac{1}{2} - \epsilon) core(v) - H\epsilon, (2 + \epsilon) core(v) + H\epsilon]$,
        \item if $f(v) \geq H$, then $core(v) \geq (\frac{1}{2} - \epsilon) H$.
    \end{itemize}
    The algorithm takes: 
\begin{itemize}
    \item for initialization from an empty graph with $n$ vertices, $O(1)$ work, and $O(1)$ worst-case depth,
    \item for any batch of edge insertions, $O(\epsilon^{-14} \log^8 n)$ work per inserted edge, and $O(\epsilon^{-12} \log^7 n)$ worst-case depth for the entire batch,
    \item for any batch of edge deletions, $O(\epsilon^{-12} \log^7 n)$ work per deleted edge, and $O(\epsilon^{-10} \log^6 n)$ worst-case depth for the entire batch.
\end{itemize}
\end{theorem}

\begin{theorem}\label{thm:density-fixed}
    There is a randomized parallel batch-dynamic data structure, which, given a parameter $H \geq 1$ and $\epsilon \in (0, 0.1)$, returns either of the following:
    \begin{itemize}
        \item a guarantee that $\rho(G) \leq (1 + \epsilon)H$, with an orientation where $\delta^+(v) \leq (2 + \epsilon) H$ for all $v \in V$,
        \item a guarantee that $\rho(G) \geq (1 - \epsilon)H$
    \end{itemize}
    
    The algorithm takes:
\begin{itemize}
    \item for initialization from an empty graph with $n$ vertices, $O(1)$ work, and $O(1)$ worst-case depth,
    \item for any batch of edge insertions, $O(\epsilon^{-21} \log^8 n)$ work per inserted edge, and $O(\epsilon^{-18} \log^7 n)$ worst-case depth for the entire batch,
    \item for any batch of edge deletions, $O(\epsilon^{-18} \log^7 n)$ work per deleted edge, and $O(\epsilon^{-15} \log^6 n)$ worst-case depth for the entire batch.
\end{itemize}
\end{theorem}

Intuitively, $H$ is a parameter that hints that the graph has a maximum arboricity of $H$. If we can guarantee that $H$ is greater but still always within a constant factor of our desired measures (in $[core(v), 100 core(v)]$, for example) throughout all the stages of our dynamic algorithm, this subsection is enough to provide a $(4+\epsilon)$-factor approximation to the graph denseness measures. If $H$ is smaller than the measure of interest, the algorithm will only know that the measure is greater than $O(H)$. If $H$ is larger than the measure of interest, the algorithm will return an estimate with an error proportional to $\epsilon H$ instead of $\epsilon$ times the measure. In a later subsection, we remove the dependency on $H$ and provide a true constant factor approximation algorithm.

To prove the result, we use our batch-dynamic data structure for the $H$-balanced orientation and the relations in \cref{sec:balanced_orientations} to argue that the data structure maintains our graph density measures. Recall the following lemma, which shows the relation between the $H$-balanced orientation and the coreness.

\maincore*

If we have $H = \Theta(\frac{\log n}{\epsilon^2})$, this lemma shows that the outdegree of each vertex is a good approximation if its coreness, and we can directly use the algorithm of \cref{thm:balanced-ds} to maintain the coreness. On the other hand, the approach would be problematic for other values of $H$. If $H < \Theta(\frac{\log n}{\epsilon^2})$, the approach suffers from high error as $\epsilon H < \frac{2 \log n}{\epsilon}$. If $H > \Theta(\frac{\log n}{\epsilon^2})$ (specifically, if $H$ is polynomial in $n$), the approach suffers from inefficiency as each update in \cref{thm:balanced-ds} takes time polynomial in $H$. 

Let $B = \frac{c \log n}{\epsilon^2}$ for a sufficiently large constant $c > 0$. We resolve these two issues with separate methods, beginning with the case $H \leq B$. For this case, instead of maintaining an $H$-balanced orientation, we maintain the $B$-balanced orientation over a graph where each edge is duplicated for $\lceil \frac{B}{H} \rceil$ times. The following lemma will show how such a modification helps achieve a better error term.

\begin{lemma}\label{lem:coreblowup}
    For any undirected graph $G = (V, E)$, a positive integer $k$, and the vertex $v \in V$ with coreness $core(G, v) = t$, consider a graph $G^\prime = (V, E^\prime)$ with the same vertex set and let $E^\prime$ be the multiset where each edge in $E$ is duplicated by $k$ times. Then we have $core(G^\prime, v) = kt$.
\end{lemma}
\begin{proof}
    We have $core(G^\prime, v) \leq kt$, since we can provide the exact same permutation with $core(G, v) = t$. Suppose that $core(G^\prime, v) < kt$. We can see that $core(G^\prime, v)$ is a multiple of $k$. By providing the permutation from $core(G^\prime, v)$, we have a certificate that $core(G, v) \leq \frac{core(G^\prime, v)}{k} < t$, which is a contradiction.
\end{proof}

We slightly modify the statement of \cref{thm:balanced-ds} to account for our needs of duplicate edges.

\begin{corollary}\label{cor:balanced-ds}
    There is a deterministic parallel batch-dynamic data structure $\textsc{Balanced}(H, K)$ which, given a parameter $\epsilon \in (0,1), H, K$ where $K$ is a positive integer, maintains a $KH$-balanced orientation over a graph where each edges are duplicated by $K$ times. Note that a single edge insertion increases the number of edges in a data structure by $K$ and vice versa.
    The algorithm takes 
\begin{itemize}
    \item for initialization from an empty graph with $n$ vertices, $O(1)$ work, and $O(1)$ worst-case depth,
    \item for any batch of edge insertions, $O(K^7 H^6 \log n)$ work per inserted edge, and $O(K^6 H^6 \log n)$ worst-case depth for the entire batch,
    \item for any batch of edge deletions, $O(K^6 H^5 \log n)$ work per deleted edge, and $O(K^5 H^5 \log n)$ worst-case depth for the entire batch.
\end{itemize}
\end{corollary}

We are ready to demonstrate the proof of the case $H \le B$.

\begin{proof}[Proof of \cref{thm:kcore-fixed} for case $H \le B$]
Let $B^{\prime} = H \lceil \frac{B}{H} \rceil$. We have $B^\prime \in [B, 2B]$. Let $K = \frac{B^\prime}{H}$. We use the data structure $\textsc{Balanced}((1 + \epsilon)H, K)$ of \cref{cor:balanced-ds}. The data structure will maintain a balanced orientation. We can infer the outdegree of each vertex $\delta^+(v)$ in constant time from the data structure. 

By \cref{lem:maincore1}, \cref{lem:maincore2}, and \cref{lem:coreblowup}, we have:

\begin{itemize}
    \item if $\delta^+(v) < (1 + \epsilon) KH - \frac{2 \log n}{\epsilon}$, $(\frac{1}{2} - \epsilon) core(v) K - \frac{2 \log n}{\epsilon} \leq \delta^+(v) \leq (2+\epsilon) core(v) K + \frac{2 \log n}{\epsilon}$,
    \item otherwise, $(1 + \epsilon)KH  \leq (2 + \epsilon) core(v) K + \frac{4 \log n}{\epsilon}$.
\end{itemize}

Since $B$ is sufficiently large, we can assume $\epsilon KH \geq \frac{4 \log n}{\epsilon}$. Using this fact, the above items are equivalent to the following:

\begin{itemize}
    \item if $\frac{\delta^+(v)}{K} < H$, $\frac{\delta^+(v)}{K} \in [(\frac{1}{2} - \epsilon) core(v) - H \epsilon, (2+\epsilon) core(v) + H \epsilon]$
    \item otherwise, $H  \leq (2 + \epsilon) core(v) + H \epsilon$.
\end{itemize}

We set $f(v) = \frac{\delta^+(v)}{K}$. Then, the theorem statement is obtained by rearranging and scaling down $\epsilon$ by an appropriate constant factor. The work and depth bound follow from \cref{cor:balanced-ds}.
\end{proof}

For the case of $H > B$, using \cref{thm:balanced-ds} directly is inefficient. We sample edges randomly to balance the tradeoff between approximation error and efficiency. Specifically, we independently sample each edge with probability $p = \frac{B}{H}$. If the coreness and other measures are about $p$ times their original values, we can easily recover the original values by multiplying them by $\frac{1}{p}$. This is true, and the formal statement and proof are given in \cref{app:conc}. Using this, we can establish the proof of \cref{thm:kcore-fixed} for case $H > B$.

\begin{proof}[Proof of \cref{thm:kcore-fixed} for case $H > B$]    
We initialize a data structure $\textsc{Balanced}(B)$ from \cref{thm:balanced-ds}. We also keep a BST of the set of edges, along with the label denoting whether it is sampled. Let $p = \frac{B}{H}$. For the inserted edges, we sample them independently with probability $p$ and add the sampled ones into $\textsc{Balanced}(B)$. For the deleted edges, we look them up in the BST and remove the ones that were sampled from $\textsc{Balanced}(B)$. We then have:

\begin{itemize}
    \item if $\delta^+(v) < B$, $core(G_p, v) \in [(\frac{1}{2} - \epsilon) \delta^+(v) - O(\frac{\log n}{\epsilon}), (2+ \epsilon) \delta^+(v) + O(\frac{\log n}{\epsilon})]$,
    \item if $\delta^+(v) \geq B$, $core(G_p, v) \geq (\frac{1}{2} - \epsilon) H -O(\frac{\log n}{\epsilon})$
\end{itemize}

\fullOnly{By \cref{lem:conc-kcore1} and \cref{lem:conc-kcore2}, we have}
\shortOnly{In the full version of this paper, we show two small lemmas that imply the following:}
\begin{itemize}
    \item if $\frac{H}{B} \delta^+(v) < H$, $core(G, v) \in [(\frac{1}{2} - O(\epsilon)) \frac{H}{B} \delta^+(v) - H  \cdot O(\epsilon), (2+ O(\epsilon)) \frac{H}{B}\delta^+(v) + H \cdot O(\epsilon)]$,
    \item if $\frac{H}{B}\delta^+(v) \geq H$, $core(G, v) \geq (\frac{1}{2} - O(\epsilon)) H -H \cdot  O(\epsilon)$.
\end{itemize}

We set $f(v) = \frac{H}{B} \delta^+(v)$. Then, the theorem statement can be obtained by rearranging and scaling down $\epsilon$ by a constant factor. The work and depth bound follow from \cref{thm:balanced-ds}, actually a looser bound. In the average-case term, it is even looser for a larger value of $H$ since most edges in updates will be ignored.
\end{proof}

The high-level idea for proving \cref{thm:density-fixed} is similar to \cref{thm:kcore-fixed}. However, as we need to maintain an orientation, we cannot ignore the unsampled edges as we did earlier. Thus, instead of randomly sampling a part of edges, we randomly partition edges into a smaller set and maintain each set in a way similar to \cref{thm:kcore-fixed}.

\begin{proof}[Proof of \cref{thm:density-fixed}]
First, consider the case where $H \geq \epsilon^{-1} B$. We start by replacing $H$ to $H = B \lceil \frac{H}{B} \rceil$ so that $\frac{H}{B}$ is an integer and $H$ is increased by at most $\epsilon H$. We create $T = \frac{H}{B}$ \textit{buckets}, where each bucket is a data structure $\textsc{Balanced}(B)$ from \cref{thm:balanced-ds}. Each edge will be put on one of the randomly selected buckets: Each bucket contains a set of edges sampled independently with probability $\frac{1}{T}$. We additionally keep the BST of a set of edges to maintain which edges belong to which partitions. For the inserted edges, we put each edge in a random bucket. For the deleted edges, we look them up in the BST and remove them from the respective buckets. To prevent spending $O(\frac{H}{B})$ time on initializing each bucket, we use a lazy initialization as done in \cref{lem:lazy_init}.

Recall that for each bucket, we report the outdegree of each vertex as its estimate $f(v)$ of its coreness. Let $\delta^+_i(v)$ be the outdegree of vertex $v$ in the $i$-th bucket. If we have $\delta_i^+(v) < B$ for all buckets, then we take the union of all buckets as an orientation. As $\sum \delta_i^+(v) < B \frac{H}{B} \leq H$, we are done, and this orientation automatically guarantees that $\rho(S) \leq H$ by \cref{lem:density}. The perturbation to make $\frac{H}{B}$ integer accounts for the $(1+ \epsilon)$ error.

Otherwise, there is a bucket $i$ and a vertex $v$ with $\delta_i^+(v) \geq B$. This bucket maintains a $B$-balanced orientation of a sampled graph $G_{p, i} = (V, E_{p, i})$. We obtain a subset of edges $G^\prime_{p, i} = (V, E^\prime_p)$ where $G^\prime_{p, i}$ admits a balanced orientation, by arbitrarily removing $\max(0, \delta^+_i(v) - B)$ out degrees of each vertex $v$: Removing such edges do not violate a $B$-balancedness, and if $\delta^+_i(v) \leq B$ holds for all $v$ the orientation is balanced.

Given that we have a balanced orientation with $\max_{v \in V} \delta^+_i(v) = B$, we have the following properties with high probability:
\begin{align*}
B = \max_{v \in V} \delta_i^+(v) \leq (1 + \frac{\epsilon}{2}) \rho(G^\prime_{p, i}) + \frac{4 \log n}{\epsilon} \leq(1 + \epsilon) \rho(G_{p, i}) + B \cdot O(\epsilon) \tag*{(\cref{lem:density})} \\
B \leq (1 + \epsilon) ((1 + \epsilon) \frac{1}{T} \rho(G) + O(\frac{\log n}{\epsilon})) + B \cdot O(\epsilon) \tag*{(\cref{lem:conc-dsg})} \\
B \leq (1 + O(\epsilon))\frac{B}{H} \rho(G) + B \cdot O(\epsilon)\\
(1-O(\epsilon)) H \leq \rho(G).
\end{align*}

Consider the other case where $H < \epsilon^{-1} B$. Let $B^{\prime} = \epsilon H \lceil \frac{B}{\epsilon H} \rceil$. We have $B^\prime \in [B, 2B]$. Let $K = \frac{B^\prime}{\epsilon H}$, which is an integer. We create a data structure $\textsc{Balanced}(H, K)$ from \cref{cor:balanced-ds}, which will maintain a $HK$-balanced orientation of $G$ with each edge duplicated by $K$ times - we denote this as $G^K$. 

Assume that $\delta^+(v) < H K$ for all $v$. Observe that $\rho(G^K) = K \rho(G)$. Hence, we obtain $K \rho(G) \leq \max_{v \in V} \delta^+(v) \leq H K$ by \cref{lem:density}, which guarantees that $\rho(G) \leq H$. To obtain an orientation of $G$ from $G^K$, for each edge, we take a direction that is assigned by the majority of duplicates in $G^K$. Then, each assignment saturates at least $K/2$ out edges of $G^K$, and we obtain an orientation where $\delta^+(v) \leq 2H$.

Otherwise, using an identical procedure of removing arbitrary outedges above, we obtain a balanced orientation for a subset of edges $G^{K\prime}$ where $\max_{v \in V} \delta^+(v) = H K = \epsilon^{-1} B^\prime$. By \cref{lem:density}, we obtain 
\begin{align*}
     \epsilon^{-1} B^\prime \leq \max_{v \in V} \delta^+(v) \leq (1 + \frac{\epsilon}{2}) \rho(G^{K\prime}) + \frac{4 \log n}{\epsilon} \\
    (1- O(\epsilon)) \epsilon^{-1} B^\prime \leq \rho(G^{K\prime}) \leq \rho(G^K) \leq K \rho(G) \\
    (1- O(\epsilon)) \epsilon^{-1} B^\prime\leq \frac{ \epsilon^{-1} B^\prime}{H} \rho(G) \\
    (1- O(\epsilon)) H \leq \rho(G) \\
\end{align*}

Finally, all $O(\epsilon)$ terms can be adjusted to $\epsilon$ by modifying $\epsilon$ by a constant factor. 

The work and depth bound follow from \cref{cor:balanced-ds}, where $H < B \epsilon^{-1}$ case dominates.
\end{proof}

\begin{remark}
    In the case where $H \geq O(\frac{\log n}{\epsilon^3})$, we indeed obtain an orientation $\delta^+(v) \leq (1 + \epsilon) H$. Our guarantee of $\delta^+(v) \leq (2 + \epsilon) H$ is only necessary when the outdegree is very small.
\end{remark}

\subsection{General algorithms, without assuming arboricity upper bounds}\label{sec:kcore_all}
By maintaining the data structure of \cref{thm:kcore-fixed} and \cref{thm:density-fixed} for each power of $(1 + \epsilon)$, we can maintain all the denseness measures unconditionally without assuming a good arboricity upper bound $H$. We first begin with the result for the coreness.

\kcore*

\begin{proof}    
    We run the algorithm of \cref{thm:kcore-fixed} for each $H_i = (1 + \epsilon)^i$ for each nonnegative integer $0 \le i \le O(\frac{\log n}{\epsilon})$. Let $f_i(v)$ be the coreness estimate $f$ maintained by the $i$-th data structure. Consider the first $k$ such that $f_k(v) < H_k$, which certainly exists. If $k = 0$, we conclude that $core(v) \leq 2$. Otherwise, we have $f_k(v) < H_k$ and $f_{k-1}(v) \geq H_{k-1}$. It follows that:
    \begin{align*}
        core(v) \leq (2 + \epsilon) (1 + \epsilon)^k + (1 + \epsilon)^k \epsilon = (2 + O(\epsilon)) (1 + \epsilon)^k \\
        core(v) \geq (\frac{1}{2} - \epsilon) (1 + \epsilon)^{k-1} - (1 + \epsilon)^{k-1} \epsilon = (\frac{1}{2} - O(\epsilon)) (1 + \epsilon)^k
    \end{align*}
    We declare $core_{ALG}(v) = (1 + \epsilon)^k$. Then, the theorem statement can be obtained by rearranging and scaling down $\epsilon$ by an appropriate constant factor. The work and depth bound follow from \cref{thm:kcore-fixed}, with an additional $O(\frac{\log n}{\epsilon})$ term for running each data structure in parallel.
\end{proof}

Now we show \cref{thm:density}, with an identical proof outline to the one above. Note that for every orientation, there exists a vertex $v$ where $\delta^+(v) \geq \rho(G)$, by the same reason from \cref{lem:density}.
\density*
\begin{proof}    
    We run the algorithm of \cref{thm:density-fixed} for each $H_i = (1 + \epsilon)^i$ for each nonnegative integer $0 \le i \le O(\frac{\log n}{\epsilon})$. Consider the first $k$ such that the $k$-th data structure guarantees $\rho(G) \leq (1 + \epsilon) H_k$, which certainly exists. If $k = 0$, we conclude that $\rho_{ALG} \leq (1 + \epsilon)$. Otherwise, we have $\rho(G) \in [(1 + \epsilon)^{k-1} ( 1- \epsilon), (1 + \epsilon)^{k+1}] = [(1 - O(\epsilon)) (1 + \epsilon)^k, (1 +O(\epsilon)) (1 + \epsilon)^k]$.
    
    For the output estimate $\rho_{ALG}$ of density, we declare $\rho_{ALG} = (1 + \epsilon)^k$. Additionally, the $k$-th data structure will maintain an orientation where $\delta^+(v) \leq (2 + \epsilon) (1 + \epsilon)^k \leq (2 + O(\epsilon)) \rho(G)$ for all $v \in V$. Then, we can obtain the theorem statement by rearranging and scaling down $\epsilon$ by an appropriate constant factor. The work and depth bound follow from \cref{thm:density-fixed}, with an additional $O(\frac{\log n}{\epsilon})$ term for running each data structure in parallel.

    For the output estimate $\lambda_{ALG}$ of arboricity, we set $\lambda_{ALG} = 2\rho_{ALG}$. Since $\rho(G) \leq \lambda(G) \leq 2 \rho(G)$ (see \cref{cor:arboricity}), the result follows.
\end{proof}

\section{Applications}\label{sec:application}
In this section, we use the orientation algorithms described in the previous section to derive our maximal matching and coloring algorithms.

\subsection{Data Structure}
Before stating the results, we clarify the interfaces of our low out-degree orientation data structure. In our applications, our low out-degree orientation data structure must list outgoing edges for each vertex. Moreover, after every update, our data structure needs to inform us of the updates incurred in our orientation, specifically the edges reversed in the update. Our data structure naturally supports these operations, which we clarify in the following lemma.

\begin{lemma}\label{lem:ds-lowoutdegree}
    There is a randomized parallel batch-dynamic data structure $\textsc{LowOutDegree}(H, \epsilon)$, which, given a parameter $H \geq 1$ and $\epsilon \in (0, 0.1)$, returns either of the following:
    \begin{itemize}
        \item a guarantee that $\rho(G) \leq (1 + \epsilon)H$, with an orientation where $\delta^+(v) \leq (2 + \epsilon) H$ for all $v \in V$,
        \item a guarantee that $\rho(G) \geq (1 - \epsilon)H$.
    \end{itemize}
    Specifically, if the data structure returned a guarantee that $\rho(G) \leq (1 + \epsilon)H$ and the orientation, that orientation supports the following interfaces:
    \begin{itemize}
        \item Given a vertex $v \in V$, the orientation can return a hash table $\mathcal{D}_{out}(v)$ that stores the set of outgoing edges in the orientation.
        \item After each edge insertion updates of size $U$, the orientation returns a hash table $\mathcal{D}_{ins}$ of size at most $O(U \epsilon^{-18} \log^6 n)$ that stores the set of possibly updated edges and its new orientation.
        \item After each edge deletion updates of size $U$, the orientation returns a hash table $\mathcal{D}_{del}$ of size at most $O(U \epsilon^{-15} \log^5 n)$ that stores the set of possibly updated edges along with its new orientation.
    \end{itemize}

    The algorithm takes 
    \begin{itemize}
    \item for initialization from an empty graph with $n$ vertices, $O(1)$ work, and $O(1)$ worst-case depth,
    \item for any batch of edge insertions, $O(\epsilon^{-21} \log^8 n)$ work per inserted edge, and $O(\epsilon^{-18} \log^7 n)$ worst-case depth for the entire batch,
    \item for any batch of edge deletions, $O(\epsilon^{-18} \log^7 n)$ work per deleted edge, and $O(\epsilon^{-15} \log^6 n)$ worst-case depth for the entire batch.
    \item for any interface access, $O(1)$ work and $O(1)$ worst-case depth.
        \end{itemize}
\end{lemma}\begin{proof}
    We reiterate the proof of \cref{thm:density-fixed} and amend our algorithm to obtain such an estimate. Again, we set $B = \frac{c\log n }{\epsilon^2}$ for a sufficiently large constant $c > 0$.

    Consider the case where $H \geq \epsilon^{-1}B$. For each vertex $v$, we initialize $\mathcal{D}_{out}(v)$ lazily, as we've done in \cref{lem:lazy_init} (but using a hash table instead of a BST). Recall that we created $T = \frac{H}{B}$ instances of data structure $\textsc{Balanced}(B)$, where each copy holds a partition of our edges. In each $\textsc{Balanced}(B)$, edges are oriented exactly as we want our low-outdegree orientation to be, and each vertex stores its outgoing edges in a binary search tree (BST). Hence, $\mathcal{D}_{out}(v)$ needs to store the union of such BSTs across the partition. For this, whenever we are performing any update in the BST of outgoing edges, we also perform the same updates to the corresponding hash table of $\mathcal{D}_{out}(v)$. This does not increase our asymptotic complexity.

    In our data structure of \cref{thm:balanced-ds}, we perform every batch edge reversal using \cref{lem:adj_rev}, which takes $O(H \log n)$ depth and work per edge. For each of these reversals, we can keep the list of reversed edges and fetch them into either $\mathcal{D}_{ins}$ or $\mathcal{D}_{del}$ in $O(\log^* n)$ depth and $O(1)$ work per edge. 

    Consider the other case where $H < \epsilon^{-1} B$. We assume $K$ to be odd; otherwise, we can increase it by $1$. We first assume that each edge in the graph is assigned distinct indices, which can be assigned to each edge when insertion updates happen. For each vertex $v$, we initialize $\mathcal{D}_{out}(v)$ lazily (but using a hash table instead of a BST), as we've done in \cref{lem:lazy_init}. Recall that we create a single instance of data structure $\textsc{Balanced}(H, K)$ in the proof. Our data structure $\mathcal{D}_{out}(v)$ should only contain the set of edges, wherein the BST of outgoing edges in $\textsc{Balanced}(H, K)$, there are at least $K/2$ copies aligned in the same direction. We additionally initialize the hash table $\mathcal{D}_{count}$, which takes the edge indices as a key and stores the number of copies oriented toward the vertex with a higher index number. For example, consider the edge $e = (u, v)$ with $u < v$ and an index $i$. Then $e$ should be oriented from $u$ to $v$ if and only if $\mathcal{D}_{count}(i) > \frac{K}{2}$.
    
    In our algorithm, every insertion and deletions are performed by \cref{lem:adj_update}, and every reversal is performed by \cref{lem:adj_rev}. Each operation will change the number of copies oriented toward the vertex with a higher index number. We store all such changes and use them to update $\mathcal{D}_{count}$ accordingly. Finally, for every update in $\mathcal{D}_{count}$, we keep the changes that flip the actual orientations (in other words, list of edges which has its count moved between $\frac{K}{2}$), and fetch them into $\mathcal{D}_{out}(\cdot), \mathcal{D}_{ins}, \mathcal{D}_{del}$ accordingly. This does not increase our asymptotic complexity.

    In our data structure of \cref{thm:balanced-ds}, we perform every batch edge reversal using \cref{lem:adj_rev}. We amend these reversal procedures to keep the list of reversed edges. Then we fetch them into either $\mathcal{D}_{ins}$ or $\mathcal{D}_{del}$ in $O(\log^* n)$ depth and $O(1)$ work per edge. 

    From the work bound of \cref{cor:balanced-ds}, we obtain a size bound of $O(\epsilon^{-18} \log^6 n)$ per each inserted edge and $O(\epsilon^{-15} \log^5 n)$ per each deleted edge. The discrepancy between the work and this size results from the fact that \cref{lem:adj_rev} only generates $O(1)$ entries per edge update while using $O(H \log n)$ work.
\end{proof}

\subsection{Maximal Matching}
In this section, we prove the following:

\matching*
\begin{proof}[Proof of \cref{thm:matching}]
We use the data structure from \cref{lem:ds-lowoutdegree}, $\textsc{LowOutDegree}(1.1 \rho_{max}, 0.05)$. As we've assumed our graph to have its density bounded by $\rho_{max}$, the data structure will never return a guarantee that $\rho(G) \geq (1 - 0.05) 1.1 \rho_{max}$. Hence, the data structure will maintain an orientation where $\delta^+(v) \leq 3 \rho_{max}$ for all $v \in V$. 

Our algorithm will additionally maintain the following data structures as a hash table: 
\begin{itemize}
    \item A single hash table $\mathcal{D}_{used}$, which maintains the set of vertices that are endpoints of any edge in the current maximal matching and 
    \item A single hash table $\mathcal{D}_{match}$, which maintains the set of edges in the maximal matching and
    \item $n$ hash tables $\mathcal{D}_{incoming}(v)$, which maintains the set of vertices that is not an endpoint of any matching and has an edge oriented toward the vertex $v$, where the orientation is consistent with $\mathcal{L}$.
\end{itemize}
For each entries of $\mathcal{D}_{ins}$ and $\mathcal{D}_{del}$, we need to update at most two elements in $\mathcal{D}_{incoming}(\cdot)$, with $O(\log n)$ depth and work. As their sizes are bounded by $O(\log^6 n)$ per edge, we conclude that the whole data structure can be maintained in $O(\log n)$ worst-case depth and $O(\log^7 n)$ work, which is well under our requirement.

When the set of edges is added or removed from our current maximal matching, we need to update each of the above data structures. Maintaining $\mathcal{D}_{used}, \mathcal{D}_{match}$ is straightforward and can be done with $O(\log n)$ worst-case depth and $O(\log n)$ work for each edge. To maintain $\mathcal{D}_{incoming}(\cdot)$, we need to iterate all vertices that are either inserted or deleted in $\mathcal{D}_{used}$, and either insert or delete their outgoing edges, according to the current orientation $\mathcal{D}_{out}(\cdot)$. This can be done in $O(\log n)$ worst-case depth and $O(\rho_{max})$ work per edge in our update.

The rest of the proof follows from the result of \cite{liu2023parallelbatchdynamicalgorithmskcore} since we've supplemented all the required data structures in their algorithm. Note that although their algorithm is stated in a way that it's amortized with high probability, it can be verified that their algorithm has a worst-case guarantee as long as the low-outdegree orientation can be maintained in the worst-case.
\end{proof}

\subsection{Explicit Coloring}
In this section, we prove the following:
\explicit*

\begin{proof}[Proof of \Cref{thm:explicit}] We use the data structure from \cref{lem:ds-lowoutdegree}, $\textsc{LowOutDegree}(1.1 \rho_{max}, 0.05)$. As we've assumed our graph to have its density bounded by $\rho_{max}$, the data structure will never return a guarantee that $\rho(G) \geq (1 - 0.05) 1.1 \rho_{max}$. Hence, the data structure will maintain an orientation where $\delta^+(v) \leq 3 \rho_{max}$ for all $v \in V$. 

Let $C = 300\rho_{max} \log n$. For each of the $n$ vertices, we initialize a \textit{palette} $c(v)$ by adding each color in $\{1, 2, \ldots, C\}$ to $c(v)$ with probability $\frac{1}{2\rho_{max}}$, independently. Notice that, with high probability, we have $|c(v)|=\Theta(\log n)$ for each node $v$. The \textit{palette} is chosen at the beginning of the algorithm and will be fixed throughout the dynamic algorithm's entire (polynomial-time) run of the dynamic algorithm. In other words, an edge update will not affect the palette of any vertices. Each vertex will store those palettes in a hash table. 

Given an low-outdegree orientation, let $\mathcal{D}_{out}(v) = \{w_1, w_2, \ldots, w_k\}$ be the set of out-neighbors of $v$. An explicit coloring is obtained in the following way: For each vertex $v$, we choose any color from $c(v) \setminus (\cup_{k} c(w_i))$ and declare it as the color of vertex $v$. One can see that, with high probability, $c(v) \setminus (\cup_{k} c(w_i))$ is not empty, as we argue next: For each color, the probability that it is ``good" in the sense that it is in $c(v)$ and not in $(\cup_{k} c(w_i)$ is at least $\frac{1}{2\rho_{max}} (1-\frac{1}{2\rho_{max}})^{k} \geq \frac{1}{2\rho_{max}} 4^{-\frac{k}{2\rho_{max}}} \geq  \frac{1}{2\rho_{max}} 4^{-1.5}> \frac{1}{20\rho_{max}}$. Now, among the $C = 300\rho_{max} \log n$, the probability that none of them is good is at most $(1-\frac{1}{20\rho_{max}})^{300\rho_{max} \log n} \leq e^{-15\log n}\leq n^{-15}$. As a result, with high probability, we can choose a color from $c(v) \setminus (\cup_{k} c(w_i))$, and one can observe that this is a valid coloring.

We now focus on recomputing the colors when the updates are given to the graph. When the set of edges is added or removed in our graph, we need to assign a new color for each vertex $v$ that has the set $\mathcal{D}_{out}(v)$ updated. To find the color for a vertex $v$, we need to try each of the colors in the palette $c(v)$ and check if it is not in any of the $c(w_i)$: For each color, we can do this in $O(\log n)$ depth and $O(\rho_{max})$ work. Trying this for all possible colors in the palette $c(v)$ takes $O(\log^2 n)$ depth and $O(\rho_{max} \log n)$ work. For each inserted edge, there are at most $O(\log^6 n)$ vertices where the set $\mathcal{D}_{out}(v)$ is updated. As we can iterate everything in parallel, the depth remains to be $O(\log^2 n)$, and we need to do $O(\rho_{max} \log^7 n)$ work per each inserted edge. The same analysis gives $O(\rho_{max} \log^6 n)$ additional work per each deleted edge.

Our final issue is that the initialization of the palettes takes $O(\rho_{max} \log n)$ work per vertex, which is quite heavy. For this, we defer the initialization of each palette to the time when the vertex first acquires any incident edges. Then, the palette initialization cost can be charged to the edge insertion, and we can use \cref{lem:lazy_init} to initialize empty palettes. 
\end{proof}

\subsection{Implicit Coloring}
In this section, we prove the following:
\implicit*
\begin{proof}[Proof of \cref{thm:implicit}]
We begin by obtaining the implicit $2^{O(\rho(G))}$-coloring.  Let $\epsilon = 0.05$, and $H_i = (1 + \epsilon)^i$ for each nonnegative integer $0 \le i \le l-1$ where $l = O(\frac{\log n}{\epsilon})$. Similarly as in \cref{thm:density}, we initialize the data structure $\textsc{LowOutDegree}(H_i, \epsilon)$ for each $0 \le i \le l-1$. However, instead of using hash table $\mathcal{D}_{out}(v)$ for maintaining the outgoing edges, we use a binary search tree to maintain the set of outgoing edges in a way that each outgoing edges are ordered in an increasing order of edge indices (which we assign in each insertion updates). We denote this binary search tree as $\mathcal{B}_{out}(v)$. Note that this can be done without changing any of the lemma statements. 

For each $i$ where the orientation of $\delta^+(v) \leq (2 + \epsilon) H_i$ is available, we maintain an implicit decomposition of the graph in the following way: For each $j = 1, 2, \ldots, \lfloor (2 + \epsilon) H_i \rfloor$, we define $F_{i, j}$ to be a directed graph with a same vertex set and the following edge set: For each edge $(v \rightarrow w)$ in the orientation $\textsc{LowOutDegree}(H_i, \epsilon)$, if the edge is the $k$-th edge in the data structure $\mathcal{B}_{out}(v)$, we put the edges into the graph $F_{i, k}$. In this way, $F_{i, j}$ forms a partition of all edges of the graph, and each of the $F_{i, j}$ forms a directed graph where each vertex has at most one outgoing edge. 

For a vertex $v$, let $w$ be a \textit{successor} of $v$ in $F_{i, j}$ if the only outgoing edge of $v$ in $F_{i, j}$ leads to $w$. Note that a successor of $v$ in $F_{i, j}$ can be computed in $O(\log n)$ time within the current data structure since we can query the $j$-th smallest edge index in the binary tree of $\mathcal{B}_{out}(v)$ in $\textsc{LowOutDegree}(H_i, \epsilon)$ in $O(\log n)$ depth and work per each query. 

We are ready to show the algorithm to compute the implicit $O(2^{O(\rho(G))})$-coloring of the graph. As in the proof of \cref{thm:density}, we let $i$ be the first $i$ such that $\rho(G) \leq (1 + \epsilon)H_i$ is guaranteed. Hence, we have $\delta^+(v) \leq 3\rho(G)$ for all $v$. Given that we can compute the successor of each queried vertex in $O(\log n)$ depth and work, we can use the classic distributed algorithm of Cole and Vishkin \cite{cole1986deterministic} to compute the $3$-coloring of each $F_{i, j}$, by retrieving the $O(\log^* n)$ successors and performing a simple local computation. This takes $O(\log^* n \log n)$ depth and work per vertex. To obtain the $3^{3\rho(G)}$-coloring, we can compute the $3$-coloring for all $j$ and combine them as a base-$3$ integer of $3 \rho(G)$ digits. In total, in $O(\log^*n \log n)$ depth and $O(\log^*n \log n \rho(G))$ work for each vertex, we can obtain a $3^{3 \rho(G)} = 2^{O(\rho(G)}$-coloring. 

To improve the number of colors to $O(\rho(G)^2)$, we compute the $2^{O(\rho(G)}$-coloring for all vertices that are at most $2$ edges apart from the queried vertices: In other words, we take the queried vertices, all of its out-neighbors, all of its out-neighbors out-neighbors, and compute the $2^{O(\rho(G))}$-coloring of them. 

Christiansen et al. \cite{christiansen2023improved} pointed out the following: Given a $k$-coloring of the graph with at most $d$ outdegrees, the local recoloring scheme of Linial \cite{linial1992locality} can deterministically compute the $k^\prime = O(d^2 (\log_d k)^2)$-coloring. Linial's scheme interprets each color of vertex $c(v)$ as a polynomial of degree $O(\log_d k)$ in the field of size $O(d \log_d k)$. Then, the candidate of possible colors out of $k^\prime$ ones (called the \textit{palette}) are found by evaluating the polynomial for all $O(d \log_d k)$ elements in the field. We note that any polynomial of degree $O(\log_d k)$ can be evaluated in $O(\log_d k)$ work with $O(\log \log_d k)$ depth for adding $O(\log_d k)$ numbers, with $O(d \log_d^2 k \log \log_d k)$ initialization to precompute all possible values of $a^b$.

Applying this to our $2^{O(\rho(G))}$-coloring, we obtain a $O(\rho(G)^4)$-coloring with $O(\rho(G)^3)$ algebraic operations for each node with $O(\log \rho(G))$ depth. This amounts to $O(\rho(G)^5)$ work per each queried vertex. 

From the $O(\rho(G)^4)$-coloring, we can apply the same method again to obtain an $O(\rho(G)^2)$-coloring. The number of algebraic operations is negligible compared to the previous reductions.

The initialization and update bound follow from \cref{lem:ds-lowoutdegree}. We have two modifications to consider: One for running a $O(\frac{\log n}{\epsilon})$ copies of $\textsc{LowOutDegree}(H_i, \epsilon)$ which does not affect the depth but incurs a $O(\log n)$ factor on work, another for replacing the hash table into the binary search tree which does not affect our statement.
\end{proof}


\bibliographystyle{alpha} 
\bibliography{library}

\newpage
\appendix
\section{Concentration of Graph Denseness Measures}\label{app:conc}
In this section, we argue that by sampling each edge independently with probability $p$, the denseness measures are also reduced by the factor of $p$, with high probability, subject to some details made precise below.

\subsection{Coreness}
\begin{lemma}\label{lem:conc-kcore1}
    Given a graph $G = (V, E)$, let $G_p = (V, E_p)$ be a graph where $E_p$ is a set of edges sampled from $E$ independently with probability $p \in (0, 1)$. For each vertex $v \in V$, $core(G_p, v) \leq (1 + \epsilon) p \cdot core(G, v) + O(\frac{\log n}{\epsilon})$ with high probability.
\end{lemma}
\begin{proof}
    Consider a permutation $per$ that is a proof for $core(G, v)$. For all nodes $u \leq_{per} v$, let $X_u$ be the random variable denoting the number of edges $u <_{per} w, (u, w) \in E_p$ plus $O(\frac{\log n}{\epsilon^2})$ copies of Bernoulli variable with expectation $1$. These are the sum of independent Bernoulli variables with a mean of at least $O(\frac{\log n}{\epsilon^2})$. By Chernoff's bound (see \cref{thm:chernoff}), with high probability the number is at most $(1 + \epsilon) p \cdot core(G, v) + O(\frac{\log n}{\epsilon})$. The lemma now follows by a union bound for all $u \leq_{per} v$ and a union bound for all $v \in V$.
\end{proof}
\begin{lemma}\label{lem:conc-kcore2}
    Given a graph $G = (V, E)$, let $G_p = (V, E_p)$ be a graph where $E_p$ is a set of edges sampled from $E$ independently with probability $p \in (0, 1)$. For each vertex $v \in V$, $core(G_p, v) \geq (1 - \epsilon) p \cdot core(G, v) - O(\frac{\log n}{\epsilon})$ with high probability.
\end{lemma}
\begin{proof}
    Fix the vertex $v$. By \cref{lem:coreequiv}, there exists a set $\{v\} \subseteq S \subseteq V$ where $G[S]$ has a minimum degree of $core(G,v)$. For each vertex $u \in S$, let $X_u$ be the random variable denoting the number of edges $(u, w) \in G_p[S]$ plus $O(\frac{\log n}{\epsilon^2})$ copies of Bernoulli variable with expectation $1$. These are the sum of independent Bernoulli variables with mean at least $O(\frac{\log n}{\epsilon^2})$. By Chernoff's bound (see \cref{thm:chernoff}), w.h.p the number is at least $(1 - \epsilon) p \cdot core(G, v) - O(\frac{\log n}{\epsilon})$. By union bound, the same holds for the degree of all vertices $v \in S$, which means $G_p[S]$ has a minimum degree at least $(1 - \epsilon)p \cdot core(G,v) - O(\frac{\log n}{\epsilon})$. By \cref{lem:coreequiv} and union bound, for every $v \in V$, we have $core(G_p, v) \geq (1 - \epsilon)p \cdot core(G, v) - O(\frac{\log n}{\epsilon})$ w.h.p.
\end{proof}

\subsection{Arboricity and Densest subgraph}
\begin{lemma}\label{lem:conc-arb}
    Given a graph $G = (V, E)$, let $G_p = (V, E_p)$ be a graph where $E_p$ is a set of edges sampled from $E$ independently with probability $p \in (0, 1)$. Then, $\lambda(G_p) \in [(1 - \epsilon) p \cdot \lambda(G) - O(\frac{\log n}{\epsilon}), (1 + \epsilon) p \cdot \lambda(G) + O(\frac{\log n}{\epsilon})]$ with high probability.
\end{lemma}
\begin{proof}
    By \cref{thm:nw}, it suffices to argue that $\frac{|E_p[S]|}{|S| - 1} \in [(1-\epsilon) p \frac{|E[S]|}{|S| - 1} - O(\frac{\log n}{\epsilon}), (1 + \epsilon) p \frac{|E[S]|}{|S| - 1} + O(\frac{\log n}{\epsilon})]$ for all $S \subseteq V, |S| \geq 2$ with high probability. We argue this for a fixed $k = |S|$; from there, the statement follows by a union bound on all $2 \le k \le |V|$.

    For a fixed $S$, consider the random variable $X$ denoting the number of edges $(u, v) \in E_p[S]$ plus $\frac{c(k-1) \log n}{\epsilon^2}$ copies of Bernoulli variable with expectation $1$ for sufficiently large constant $c$. These are the sum of independent Bernoulli variables with mean at least $\frac{c(k-1) \log n}{\epsilon^2}$. By Chernoff's bound (\cref{thm:chernoff}), we have $(1 - \epsilon) (pE[S] + \frac{c(k-1) \log n}{\epsilon^2}) \le X + \frac{c(k-1) \log n}{\epsilon^2} \le (1 + \epsilon)(pE[S] + \frac{c(k-1) \log n}{\epsilon^2})$ with probability at least $1 - n^{-O(k)}$. Taking a union bound for all $\binom{n}{k} = O(n^k)$ possible sets $S$, we see that this holds for all sets with high probability, as desired. 
\end{proof}

The same argument can be shown with identical proof for the densest subgraph.
\begin{lemma}\label{lem:conc-dsg}
    Given a graph $G = (V, E)$, let $G_p = (V, E_p)$ be a graph where $E_p$ is a set of edges sampled from $E$ independently with probability $p \in (0, 1)$. Then, $\rho(G_p) \in [(1 - \epsilon) p \cdot \rho(G) - O(\frac{\log n}{\epsilon}), (1 + \epsilon) p \cdot \rho(G) + O(\frac{\log n}{\epsilon})]$ with high probability.
\end{lemma}

\end{document}

%% file: macros.tex
\usepackage[IL2]{fontenc}
\usepackage[letterpaper,margin=1.00in]{geometry}
\usepackage{amsmath, amssymb, amsthm, amsfonts}
\usepackage{bbm, dsfont}
\usepackage{amscd}
\usepackage{mathrsfs}

\usepackage{comment} 
\usepackage{ifthen}
\usepackage{tikz}
\usetikzlibrary{positioning,decorations.pathreplacing}
\usepackage{ bbold }
\usepackage{appendix}
\usepackage{graphicx}
\usepackage{color}
\usepackage{epstopdf}
\usepackage{wrapfig}
\usepackage{paralist}
\usepackage{wasysym}
\usepackage[textsize=tiny]{todonotes}

\usepackage{listings} 

\usepackage{pdflscape} 

\usepackage{framed}
\usepackage[framemethod=tikz]{mdframed}
\usepackage[bottom]{footmisc}
\usepackage{enumitem}
\setitemize{noitemsep,topsep=3pt,parsep=3pt,partopsep=3pt}
\usepackage[font=small]{caption}
\usepackage{xspace}

\usepackage{thmtools} 
\usepackage{thm-restate} 

\definecolor{darkgreen}{rgb}{0,0.5,0}
\definecolor{darkblue}{rgb}{0,0,0.6}
\usepackage{hyperref}
\hypersetup{
    unicode=false,          
    colorlinks=true,        
    linkcolor=darkblue,          
    citecolor=darkgreen,        
    filecolor=magenta,      
    urlcolor=cyan           
}

\newtheorem{theorem}{Theorem}[section]
\newtheorem{lemma}[theorem]{Lemma}
\newtheorem{meta-theorem}[theorem]{Meta-Theorem}

\newtheorem{remark}[theorem]{Remark}
\newtheorem{corollary}[theorem]{Corollary}

\newtheorem{definition}[theorem]{Definition}

\usepackage[capitalize, nameinlink,noabbrev]{cleveref}

\crefname{theorem}{Theorem}{Theorems}
\crefname{proposition}{Proposition}{Propositions}
\crefname{observation}{Observation}{Observations}
\crefname{lemma}{Lemma}{Lemmas}
\crefname{claim}{Claim}{Claims}
\crefname{problem}{Problem}{Problems}
\crefname{conjecture}{Conjecture}{Conjectures}
\crefname{question}{Question}{Questions}
\crefname{example}{Example}{Examples}
\crefname{fact}{Fact}{Facts}

\definecolor{darkgreen}{rgb}{0,0.5,0}

\usepackage{algcompatible}
\algnewcommand\algorithmicswitch{\textbf{switch}}
\algnewcommand\algorithmiccase{\textbf{case}}

\algdef{SE}[SWITCH]{Switch}{EndSwitch}[1]{\algorithmicswitch\ #1\ \algorithmicdo}{\algorithmicend\ \algorithmicswitch}%
\algdef{SE}[CASE]{Case}{EndCase}[1]{\algorithmiccase\ #1}{\algorithmicend\ \algorithmiccase}%
\algtext*{EndSwitch}%
\algtext*{EndCase}%

\newcommand{\eps}{\varepsilon}

\newcommand{\poly}{\operatorname{poly}}

\renewcommand{\phi}{\varphi}

\renewcommand{\paragraph}[1]{\vspace{0.15cm}\noindent {\bf #1}:}

\newcommand{\FullOrShort}{full}
\ifthenelse{\equal{\FullOrShort}{full}}{
  
  \newcommand{\fullOnly}[1]{#1}
  \newcommand{\shortOnly}[1]{}

  }{

    \newcommand{\fullOnly}[1]{}
    \newcommand{\IncludePictures}[1]{}
   
  }

\newcommand{\Pot}

\usepackage{algorithm}
\usepackage[noend]{algpseudocode}